\documentclass[lettersize,journal]{IEEEtran}
\usepackage{amsmath,amsfonts}
\usepackage{algorithmic}
\usepackage{subfigure}
\usepackage{array}
\usepackage[caption=false,font=normalsize,labelfont=sf,textfont=sf]{subfig}
\usepackage{textcomp}
\usepackage{stfloats}
\usepackage{url}
\usepackage{verbatim}
\usepackage{graphicx}
\def\BibTeX{{\rm B\kern-.05em{\sc i\kern-.025em b}\kern-.08em
    T\kern-.1667em\lower.7ex\hbox{E}\kern-.125emX}}
\usepackage{balance}
\usepackage{enumerate}
\usepackage{algorithm}
\usepackage{algorithmic}
\usepackage{bm}
\usepackage{booktabs}  
\usepackage{diagbox}   
\usepackage{multirow}  
\newcommand{\setParDis}{\setlength {\parskip} {0.1cm} }

\begin{document}
\title{ATASI-Net: An Efficient Sparse Reconstruction Network for Tomographic SAR Imaging with Adaptive Threshold}
\author{Muhan Wang,
	Zhe Zhang,~\IEEEmembership{Member,~IEEE,}
	Xiaolan Qiu,~\IEEEmembership{Senior Member,~IEEE,}	
	Silin Gao,
	and Yue Wang,~\IEEEmembership{Senior Member,~IEEE}
	\thanks{Part of this work was supported by the NSFC grant \#61991421, \#61991420.}
	\thanks{Part of this work was presented on the IET International Conference on Radar Systems 2022, October 24-27, 2022, Edinburgh, UK.}
	\thanks{M. Wang, Z. Zhang and X. Qiu are with Suzhou Key Laboratory of Microwave Imaging, Processing and Application Technology, and Suzhou Aerospace Information Research Institute, Suzhou, Jiangsu, China}
	\thanks{M. Wang, Z. Zhang, S. Gao and X. Qiu are also with Aerospace Information Research Institute, Chinese Academy of Sciences, and School of Electronic, Electrical and Communication Engineering, University of Chinese Academy of Sciences, Beijing, China.}
	\thanks{M. Wang, S. Gao and X. Qiu are also with the Key Laboratory of Technology in Geo-spatial Information Processing and Application System, Chinese Academy	of Sciences, Beijing, China.}
	\thanks{Y. Wang is with Electrical and Computer Engineering Department, George Mason University, Fairfax, VA, USA. }
	\thanks{Corresponding author: Zhe Zhang, Email: zhangzhe01@aircas.ac.cn.}
	\thanks{Manuscript received Nov xx, 20xx; revised August xx, 20xx.}
}

\markboth{IEEE Transactions on Geoscience and Remote Sensing,~Vol.~XX, No.~XX, September~20XX}{M. Wang \MakeLowercase{\textit{et al.}} ATASI-Net： An Efficient Sparse Reconstruction Network for Tomographic SAR Imaging with Adaptive Semantic Threshold}
\maketitle

\begin{abstract}
Tomographic SAR technique has attracted remarkable interest for its ability of three-dimensional resolving along the elevation direction via a stack of SAR images collected from different cross-track angles. The emerged compressed sensing (CS)-based algorithms have been introduced into TomoSAR considering its super-resolution ability with limited samples. However, the conventional CS-based methods suffer from several drawbacks, including weak noise resistance, high computational complexity, and complex parameter fine-tuning. Aiming at efficient TomoSAR imaging, this paper proposes a novel efficient sparse unfolding network based on the analytic learned iterative shrinkage thresholding algorithm (ALISTA) architecture with adaptive threshold, named Adaptive Threshold ALISTA-based Sparse Imaging Network (ATASI-Net). The weight matrix in each layer of ATASI-Net is pre-computed as the solution of an off-line optimization problem, leaving only two scalar parameters to be learned from data, which significantly simplifies the training stage. In addition, adaptive threshold is introduced for each azimuth-range pixel, enabling the threshold shrinkage to be not only layer-varied but also element-wise. Moreover, the final learned thresholds can be visualized and combined with the SAR image semantics for mutual feedback. Finally, extensive experiments on simulated and real data are carried out to demonstrate the effectiveness and efficiency of the proposed method.
\end{abstract}


\section{Introduction}
\IEEEPARstart{S}{ynthetic} aperture radar (SAR)~\cite{SAR} uses synthetic aperture principle to achieve high resolution in both the azimuth and range directions. However, traditional two-dimensional (2-D) SAR images do not have resolving ability along the elevation direction, causing issues such as layover and shadowing. SAR tomography (TomoSAR)~\cite{TomoSAR}, as an emerging radar imaging technique, has attracted remarkable interest in recent years for its ability in achieving three-dimensional resolution by collecting data of the same target scene from multiple slightly different viewing, by either a physical antenna array or multiple flight passes. TomoSAR technique has been widely used in urban observation~\cite{urban}, forestry detection~\cite{forest} and disaster monitoring~\cite{disaster}.

Along the elevation dimension, TomoSAR imaging problem can be treated as a line spectrum estimation problem in theory. Traditionally, it can be solved via canonical spectrum estimation algorithms, such as SVD~\cite{SVD}, MUSIC~\cite{MUSIC} or CAPON~\cite{CAPON}, which however usually experience poor performance under limited observations and low SNR circumstances. In typical scenarios of TomoSAR applications, scatters are usually distributed sparsely along the elevation direction, and meanwhile only few significant scatterers fall into a range-azimuth pixel. Thus, compressed sensing (CS)~\cite{CS} based methods are widely used to solve the TomoSAR inversion problem as the state-of-the-art approach. A myriad of techniques has been proposed to tackle such sparse inversion problems, which conventionally contain two categories. The first group is the greedy approach, adopting the   minimization, such as orthogonal matching pursuit (OMP)~\cite{OMP} and CoSaMP~\cite{CoSaMP}. The second class of methods, known as convex optimization, adopts the   regularization and forms a convex object function, such as iterative shrinkage-thresholding algorithm (ISTA)~\cite{ISTA}, approximate message passing (AMP) algorithm~\cite{AMP}, and alternating direction method of multipliers (ADMM)~\cite{ADMM}. In recent years, different CS-based methods for solving TomoSAR inversion have been extensively studied. Budillon et al.~\cite{Budillon} presented the first simulation of CS TomoSAR. X. Zhu and R. Bamler~\cite{first real} conducted the first real data tomographic SAR inversion by CS approach and developed the SL1MMER method~\cite{slimmer}. He et al.~\cite{GOMP} applied the generalized OMP algorithm to the SAR tomography imaging. Han et al.~\cite{ISTA-AT} modify the ISTA and proposed an improved method for the 3D reconstruction of airborne SAR tomography. Although CS-based algorithms are widely accepted and believed to be reliable, they still suffer from several drawbacks, including weak noise resistance, high computational complexity, the holding conditions of sparse assumption, and fine-tuning of reconstruction parameters.

With the development of artificial intelligence and deep learning technique, the data-driven deep network provides a new idea to overcome the limitation of complicated parameter fine-tuning issues in CS and improving its performance in many aspects. Thanks to the enhanced interpretability, fast implementation, and high robustness to model mismatch, a deep learning method called "deep unfolding"~\cite{deep unfold} was proposed to provide a concrete and systematic connection between iterative model-based algorithms and deep neural networks. Specifically, iteration based sparse reconstruction algorithms of CS can be represented by an unfolded deep recurrent network. The reconstruction parameters e.g. regularization parameter can be learned from a given dataset via canonical training methods over a given dataset, and the sparse reconstruction becomes a simple and fast inference process. Along this line, a novel sparse unfolding network, called LISTA, has been firstly proposed as the unfolded network of the ISTA by Gregor and LeCun~\cite{LISTA}. In LISTA, ISTA is folded to a multi-layer deep network. Following the unrolling, training data can be fed through the network, and stochastic gradient descent can be used to update and optimize its parameters, including the regularization parameter, step size, and even the measurement matrix. Borgerding and Schniter~\cite{LAMP} proposed LAMP by unfolding AMP to deep network. Unfolding was also applied to the ADMM algorithm to solve the magnetic resonance imaging (MRI) problem~\cite{ADMM-CS}. The resulting network, ADMM-CSNet, uses training data to learn filters, penalties, simple nonlinearities, and multipliers. In addition, some improved networks based on the above unfolded networks have also been proposed, including LISTA-CPSS~\cite{LISTA-CPSS}, LDAMP~\cite{LDAMP}, etc. Moving beyond natural images and medical images, the deep unfolded network has been applied successfully in remote sensing~\cite{RS1}~\cite{RS2}, speech and video processing~\cite{speech}~\cite{video}, and so on. In these applications, unfolding and training significantly improve both the quality and speed of signal reconstruction.

Recently, the idea of deep unfolding has also triggered the attention in the applications of SAR imaging. Pu et al.~\cite{wei1}~\cite{wei2} proposed a deep SAR imaging algorithm using the unfolded ADMM based auto-encoder structure  to deal with the SAR motion compensation and SAR autofoucs respectively, and eliminate the influence of motion errors, improving the quality of SAR imaging. Aiming at enhancing the desired target and improving the SCR in the reconstructed SAR images, Li et al.~\cite{Li1} unfolded an iterative ADMM solver based on matched filter methods, and proposed MF-ADMM-Net. Also based on the ADMM architecture, Li et al.~\cite{Li2} later combined the sparsity-cognizant total least-square model and proposed the STLS-LADMM-Net for improving the quality of SAR autofocus imaging. In addition, the UESTC team has extended the application of the deep unfolding framework to 3-D millimeter-wave (mmWave) SAR imaging systems. Wang et al.~\cite{TPSSI} firstly proposed a two-path iterative framework dubbed TPSSI-Net based on the AMP algorithm architecture. Rather than manually choosing a sparse dictionary, a two-path convolutional neural network is developed and embedded in TPSSI-Net for nonlinear sparse representation in the complex-valued domain. Results in~\cite{TPSSI} show that TPSSI-Net is capable of yielding favorable 3D reconstruction performance compared with traditional methods. For the same purpose of improving the performance of 3-D mmWave image reconstruction, Wei et al.~\cite{SISR} proposed SISR-Net, Zhou et al.~\cite{SAF} proposed SAF-3DNet, and Wang et al.~\cite{LFIST} proposed LFIST-Net. These deep 3D SAR imaging methods validate the superiority of deep unfolded networks through simulated and real data, however, they are all targeting the millimeter wave SAR under non-sparse scenarios.

Encouraged by deep unfolding, the TomoSAR community also started to design deep unfolded framework based on CS iterative optimization algorithms in urban areas with sparse scenes. Gao et al.~\cite{LVAMP} unfolded and mapped vector AMP~\cite{VAMP} into a deep network framework for line spectral estimation and applied it to tackle TomoSAR inversion, and Lei et al.~\cite{Lei} quantified the optimal network design and maximum SR ability of the network. Simulation results validate the superiority of the proposed deep unfolded network based on the backbone of VAMP. Qian et al.~\cite{gamma} developed a sparse unfolding network named $\gamma $-Net to solve the TomoSAR problem by improving the LISTA-CPSS. And evaluation shows that the proposed network is able to deliver competitive performance to the state-of-the-art in terms of the super-resolution capability and elevation estimation accuracy. However, although these methods can replace the traditional methods by designing the network with the idea of deep unfolding and learning the parameters, the amount parameters that are needed to be learned is usually very huge. This will lead to huge training computational complexity and enormous dataset size requirement, which is usually unacceptable in SAR applications.

In this paper, we propose a novel and efficient deep unfolded network named Adaptive Threshold ALISTA-based Semantic Imaging Network (ATASI-Net). Based on the success of analytic LISTA (ALISTA) architecture~\cite{ALISTA}, we are able to reduce the number of required training samples and increase the processing speed and improve the reconstruction accuracy. In concise, each layer of ATASI-Net consists of one module and three units in cascade, including pre-calculation module, error propagation unit, threshold unit, and reconstruction unit. Exploiting the ALISTA, in pre-calculation module of ATASI-Net, the weight matrix in LISTA is computed as the solution of an off-line optimization problem. Only the step size and threshold parameters are learned from data, which significantly simplifies the training stage. In addition to canonical ALISTA, we introduce adaptive thresholds for element-wise nonlinear operators to replace the traditional threshold shrinkage function in threshold unit. And the final learned element-wise thresholds are able to split and feed with the semantics of SAR images. Moreover, the proposed network can be trained using various training methods, and a structured approach is adopted to output the 3D point clouds.

The main contributions of this article are as follows.

\begin{itemize}
	\item [1)] 
	By combining interpretable signal processing models based iterative CS algorithm and data-driven deep networks, an efficient deep sparse unfolded network named ATASI-Net is proposed for TomoSAR imaging. The parameters in ATASI-Net can be learned from data, avoiding the parameter fine-tuning stage and reducing temporal and spatial complexity.      
	\item [2)]
	Different from LISTA, we adopt the ALISTA approach whereas the weight matrix in ATASI-Net is computed via an off-line convex optimization problem, leaving only two scalar parameters to data-driven learning. This greatly reduces the training dataset size and time consumption.
	\item [3)]
	 We adopt adaptive thresholding in the threshold module for each azimuth-range pixel, enabling the threshold shrinkage to be not only layer-varied but also element-wise. The final learned thresholds are visualized in the experiments on the measured data, and the threshold segmentation can be combined with the SAR image semantics for mutual feedback.
	 \item[4)] 
	 We use simulation data generated from prior known scene geometry parameters as the training set for the proposed network. Moreover, for real scenes with unknown geometric parameters, we propose to use the better-focused data in the results as the training set, which can improve the conventional reconstruction results. And in real data experiments, we use structured 3D modeling to evaluate the point cloud results. Finally, extensive experiments are carried out to demonstrate the superiority of the proposed ATASI-Net.
\end{itemize}
  
The remainder of the article is organized as follows. Section II introduces the TomoSAR imaging model and existing approaches. Section III proposes the details of the ATASI-Net deep network, as well as its training strategies and discussions. The simulation and real data experimental results are demonstrated in Section IV and Section V, respectively. At last, the conclusion of this paper is given in Section VI.

\section{Problem setting and Unfolded Network of ISTA}

	\subsection{TomoSAR Imaging Model}
\begin{figure}[!h]
		\centering
		\includegraphics[width=2.5in]{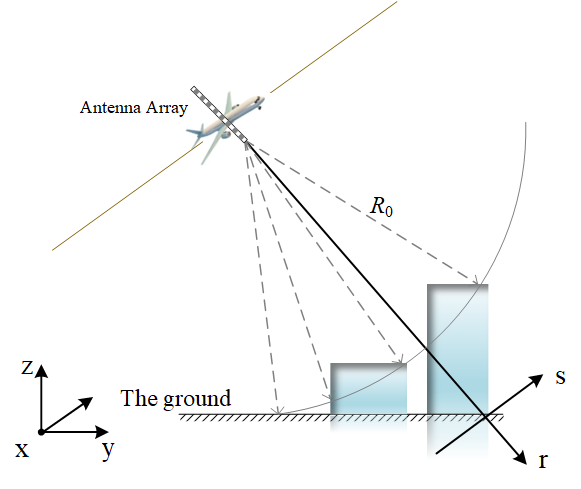}
		\caption{TomoSAR imaging model}
		\label{TomoSAR Model}
	\end{figure}

	The tomographic SAR technique uses a stack of SAR images collected from different cross-track angles of the same observation area to reconstruct the scattering information along the elevation direction and achieve the three-dimensional resolving ability. The basic principle of SAR tomography is illustrated in Fig. \ref{TomoSAR Model}.

	For a single look complex (SLC) SAR image, $y({x_0},{r_0})$ represents the value of an azimuth-range pixel $({x_0},{r_0})$. Consider the same target is observed N times from slightly different viewing angles to obtain N SLCs (via an antenna array or multiple flight passes). If the SLCs are perfectly aligned, denote the \emph{n}-th acquisition as ${y_n}$. Each azimuth-range pixel $({x_0},{r_0})$ can be expressed as the integration of the scattering coefficient along the elevation weighted by sinusoids as:

	\begin{equation}
	\label{Eq1}
		{y_n}({x_0},{r_0}) = \int {\gamma ({x_0},{r_0},s)\exp \left( {j\frac{{4\pi }}{\lambda } \cdot \frac{{s{b_n}}}{{{r_0}}}} \right)ds} 
	\end{equation}

	where $\gamma(s)$ represents the scattering coefficient distribution, a.k.a. reflectivity function along elevation s, $b_n$ is the baseline length of the \emph{n}-th observation, $\lambda$ is the wavelength of the transmitted signal. Assuming that the scatters are sparsely distributed along the elevation direction, we can discretise the continuous function $\gamma(\cdot)$ into a sparse vector. Along with the additive noise, (\ref{Eq1}) can be written in a matrix-vector form as (2).
	
	\begin{equation}
	\label{Eq2}
		{\bf{y}} = {\bf{R\bm{\gamma }}} + \bm{\varepsilon} 
	\end{equation}

	where ${\bf{y}}$ is the measurement vector stacked by $y_n$, $\bf{R}$ plays as an $N \times L$ mapping matrix with

	\begin{equation}
	\label{Eq3}
		{R_{nl}} = \exp (j4\pi {b_n}{s_l}/\lambda {r_0})
	\end{equation}

	Now, TomoSAR inversion boils down to a signal recovery problem, where the goal is to obtain the corresponding scattering parameters such as elevation and reflectivity profile $\bm{\gamma}$ of each range-azimuth cell by solving (2), where $\bf{R}$ is usually known from model and $\mathbf{y}$ is the observation.

	\subsection{Iterative Thresholding Algorithm (ISTA)}
	It is illustrated in~\cite{L1Tomo} that the number of layovered scatters, or sparsity, is less than 4 in each range-azimuth resolution cell for TomoSAR in the vast majority of urban areas. Hence, the echo signal along the elevation direction is sufficiently sparse and can be solved via a sparse reconstruction problem (2) using the compressed sensing (CS) technique. Within the CS framework, $\bm{\gamma}$ can be reconstructed by $L_0$-norm minimization:
	
	\begin{equation}
		\mathop {\min }\limits_{\bm{\gamma}}  ||{\bf{\bm{\gamma} }}|{|_0}\quad{\rm{   s}}{\rm{.t}}{\rm{.  }}\ {\bf{y}} = {\bf{R\bm{\gamma} }}
	\end{equation}
	
	and in the presence of noise, it can be approximated by:
	
	\begin{equation}
		{\bm{\hat \gamma }} = \arg \mathop {\min }\limits_{\bm{\gamma }} \frac{1}{2}||{\bf{y}} - {\mathbf{R}\bm{\gamma }}||_2^2 + \lambda ||{\bm{\gamma }}|{|_0}
	\end{equation}
	
	As a convex relaxation, $L_0$-norm can be replaced by $L_1$ regularization~\cite{L0}, and $\bm{\gamma}$ can be reconstructed in the form of least absolute shrinkage and selection operator (LASSO) problem as:

	\begin{equation}
	\label{LASSO}
	{\bm{\hat \gamma }} = \arg \mathop {\min }\limits_{\bm{\gamma }} \frac{1}{2}||{\bf{y}} - {\mathbf{R}\bm{\gamma }}||_2^2 + \lambda ||{\bm{\gamma }}|{|_1}
	\end{equation}
	
	where $\lambda$ is a regularization parameter that determined by the sparsity of $\bm{\gamma}$.
	
	Iterative soft-thresholding algorithm (ISTA) is a widely used sparse reconstruction algorithm. Compared with greedy algorithms such as OMP, ISTA has better robustness against noise and reconstruction performance, but is relatively time comsuming. In ISTA, the estimate of $\bm{\gamma}$  is achieved in an iterative manner as
	
	\begin{equation}
	\label{ISTA}
		{{\bm{\hat \gamma }}_{k + 1}} = {h_{(\alpha /L)}}({{\bm{\gamma }}_k} + \frac{1}{L}{{\bf{R}}^T}({\bf{y}} - {\bf{R}}{{\bm{\hat \gamma }}_k}))
	\end{equation}
	
	where L is a parameter controlling the iteration step size satisfying $L > {\lambda _{\max }}({{\bf{R}}^T}{\bf{R}})$, ${\lambda _{\max }}( \cdot )$ denotes the maximum eigenvalue, $\theta  = \alpha /L$ is the threshold parameter, and ${h_\theta }(X)$ is the soft-thresholding function defined as
	
	\begin{equation}
	\label{softshrinkage}
		{h_\theta }(X) = sign(X)\max (|X| - \theta ,0)
	\end{equation}
	
	However, the parameters $\alpha,L$ in the ISTA algorithm are manually chosen, which usually requires a time-consuming fine-tuning process to achieve the best performance. Furthermore, these parameters are not adaptive i.e., they are fixed from one scene to another unless we fine-tune them repeatedly. To overcome the drawbacks of traditional ISTA, LeCun et al. unfold the ISTA into a deep network to leverage the benefits of deep learning, namely Learned ISTA (LISTA) as proposed in~\cite{LISTA}.
	
	\subsection{Unfolded Deep Networks Based on ISTA}
	In LISTA, one iteration in (\ref{ISTA}) can be rewritten in a way of neuron’s activities as one layer of neural network:
	
	\begin{equation}
	\label{LISTA}
		{{\bm{\hat \gamma }}_{k + 1}} = {h_{{\theta _{k + 1}}}}\{ {\bf{W}}_1^{k + 1}{\bf{y}} + {\bf{W}}_2^{k + 1}{{\bm{\hat \gamma }}_k}\} 
	\end{equation}
	
	where ${\bf{W}}_1^{k + 1} = \frac{1}{L}{{\bf{R}}^H}$ and ${\bf{W}}_2^{k + 1} = I - \frac{1}{L}{{\bf{R}}^H}{\bf{R}}$ are two sets of parameters that can be trained. Total $K$ iterations constructs a $K$-layer neural network. It is a sparse unfolded network that takes advantage of the representation power of deep learning and uses available data to train the parameters, matrix $\bf{W}_1^k$,$\bf{W}_2^k$ and threshold $\theta_k$. Compared with ISTA, LISTA converges faster and produces a better solution as~\cite{LISTA-CPSS} demonstrated.
	
	Although LISTA allows us to learn parameters from data autonomously, the sizes of the weight matrices $\bf{W_1}$ and $\bf{W_2}$ are usually huge in real SAR signal processing applications. Hence, a vast dataset is required to train the enormous amount of parameters, which is usually impractical, and the training process is temporally and spatially complex.
	
	Tackling such challenges, we propose to introduce the Analytic LISTA (ALISTA)~\cite{ALISTA} to TomoSAR problem. ALSITA is an improved version of LISTA where both weight matrices are determined off-line via a convex optimization problem, leaving only the step size and threshold parameters to be learned. It significantly simplifies the training stage in both spatial and temporal domains. It has been shown that ALISTA retains the benefits of LISTA in terms of optimal linear convergence rate and achieves a performance comparable to LISTA.
	
	Instead of training $\bf{W_1}$,$\bf{W_2}$, weight matrix $\bf{W}$ in ALISTA is pre-determined by solving the following convex optimization problem (\ref{W}).
	
	\begin{equation}
	\label{W}
		{\bf{W}} = \mathop {\arg \min }\limits_{{\bf{W}} \in {C^{M \times N}}} ||{{\bf{W}}^T}{\bf{R}}||_F^2{\rm{s}}{\rm{.t}}{\rm{. (}}{{\bf{W}}_{:,i}}{{\rm{)}}^T}{{\bf{R}}_{:,i}} = 1,i = 1,...,n
	\end{equation}
	
	which plays as a mutual coherence minimizer between $\bf{W}$ and $\bf{R}$. This is motivated by the tenet in CS that a dictionary with smaller coherence possesses better sparse recovery performance. 
	
	Then, let ${{\bf{W}}^k} = {\mu ^k}{\bf{W}}$, (\ref{LISTA}) can be rewritten as the following:
	
	\begin{equation}
	\label{ALISTA}
		{\bm{\hat \gamma} _{k + 1}} = {h_{{\theta _k}}}\{ {\bm{\hat \gamma} _k} + {\eta _k}{{\bf{W}}^T}({\bf{y}} - {\bf{R}}{\bm{\hat \gamma} _k})\} 
	\end{equation}
	
	where only two scalar parameters $\{ {\theta _k},{\eta _k}\}  \in R$ are learned from end-to-end data. Particularly, the number of training parameters is reduced from $\mathcal{O}(KM^2+K+MN)$ in LISTA down to $\mathcal{O}(K)$ in ALISTA, causing a markedly reduction in training burden and required training dataset size.

\section{Proposed ATASI-Net for TomoSAR}
	\subsection{Adaptive Threshold}
	ALISTA still uses the traditional threshold shrinkage function where the regularization parameter is fixed in a specific layer. In practice, the scattering intensity of different target structures of SAR images often varies greatly, and taking the same threshold value for the whole SAR image will result in chopping off the information, which is inappropriate. In this work, we consider a threshold adaptation approach for sparse recovery by introducing a concave regularizer to promote sparsity. Based on this, we introduce the learnability of deep networks. In the actual data experiment, we segmented and visualized the element-wise threshold learned through the network, and verified that it can feed mutually with SAR image semantic information, which validates the rationality of the proposed adaptive thresholds.

	We formulate a sparse-promoting problem using concave regularizer G(x) to approximate the substitution of $L_0$ norm in (5).
	
	\begin{equation}
	\label{L_log}
		\mathop {{\bm{\gamma }} = \mathop {\arg \min }\limits_{\bm{\gamma }} }\limits_{} \frac{1}{2}||{\bf{y}} - {\bm{A\gamma }}||_2^2 + \lambda \sum\limits_{i = 1}^n {G(|{\gamma _i}|)} 
	\end{equation}
	
	Where, $\bf{A}$ represents the sensing matrix, which equals $\bf{R}$ in (3).
	
	~\cite{concave1}~\cite{concave2}~\cite{concave3} also used such concave regularizers for sparse recovery problems. According to~\cite{concave3}, when the derivative of $G(x)$, $g(x)=G'(x)$ satisfies non-increasing function and has $g(0) = 1$, namely, $G(x)$ is a concave increasing function. Specially, if $G(x) = x$, then it can be expressed as (\ref{LASSO}) and reduced to the LASSO problem. We can also use other different $g(x)$ such as $g(x) = \max (1 - x/\tau ,0)$~\cite{reweightedL1} and $g(x) = 1/({x^2}/{\tau ^2} + 1)$~\cite{G(x)2}.
	
\begin{figure}[!h]
	\centering
	\includegraphics[width=2.5in]{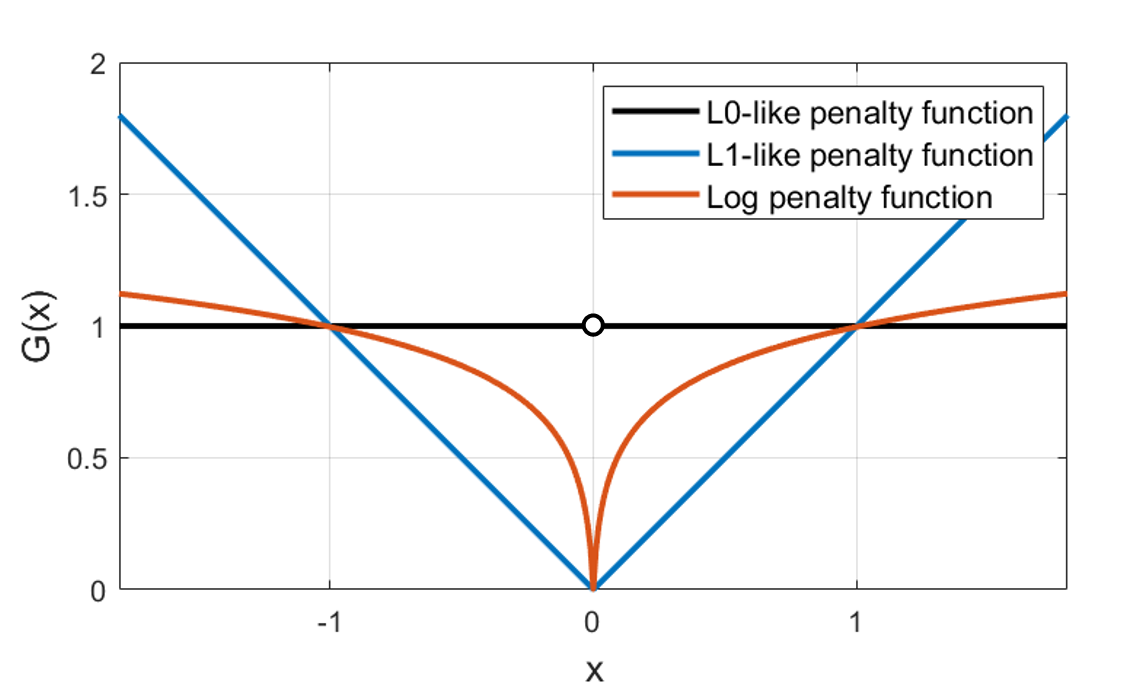}
	\caption{The canonical $L_0$ sparsity count is better approximated by the log-sum penalty function than by the traditional convex $L_1$ relaxation.}
	\label{L_logfig}
	\end{figure}

	Inspired by~\cite{reweightedL1}~\cite{Logpen} that reweighted $L_1$ minimization can improve the recovery of sparse signals. In particular, the log-sum penalty function has the potential to be much more sparsity-encouraging than the $L_1$-norm as Fig. \ref{L_logfig} shows. 
	
	In this paper, we choose strictly concave $G(x) = \tau \log (x + \tau )$, where $\tau $ is set to be a small positive number. The derivative of $G(x)$, $g(x) = G'(x) = \tau /(x + \tau )$ satisfies non-increasing function and has $g(0) = 1$. Therefore, for $G(x) = \tau \log (x + \tau )$, (\ref{L_log}) is a non-convex problem, and the strictly concave function $G(x)$ satisfies
	
	\begin{equation}
	\label{G(x)}
		G(x) \le G(y) + g(y)(x - y)
	\end{equation}
	
	And for $f({\bf{x}}) = \frac{1}{2}||{\bf{y}} - {\bf{Ax}}||_2^2$ satisfies
	
	\begin{equation}
	\label{f(x)}
		f({{\bf{x}}_1}) \le f({{\bf{x}}_2}) + \nabla f{({{\bf{x}}_2})^T}({{\bf{x}}_1} - {{\bf{x}}_2}) + \frac{1}{{2L}}||{{\bf{x}}_1} - {{\bf{x}}_2}||_2^2
	\end{equation}
	
	where $L$ is upper-bounded by the inverse of the Lipschitz constant of $\nabla f({\bf{x}})$, and $L$ satisfies $L \le 1/{\lambda _{\max }}({{\bf{A}}^T}{\bf{A}})$~\cite{Lyueshu}.
	Hence, we have
	
	\begin{equation}
	\label{jiehe}
		\begin{aligned}
		f({\bm{\gamma }}) + \lambda \sum\limits_{i = 1}^n {G(|{\gamma _i}|)}  &\le f{({{\bm{\gamma }}^k})^T}({\bm{\gamma }} - {{\bm{\gamma }}^k}) + \frac{1}{{2L}}||{\bm{\gamma }} - {{\bm{\gamma }}^k}||_2^2\\
		&+ \lambda \sum\limits_{i = 1}^n {\left( {G(|\gamma _i^k|) + g(|\gamma _i^k|)(|\gamma _i^{}| - |\gamma _i^k|)} \right)} 
		\end{aligned}
	\end{equation}
	
	~\cite{major-min} indicates that for (\ref{jiehe}), the majorization-minimization method minimizes the tight upper bound instead of minimizing the original function. Then we have
	
	\begin{equation}
	\label{tuidao_gamma}
		\begin{aligned}
		{{\bm{\gamma }}^{k + 1}} &= \mathop {\arg \min }\limits_{\bm{\gamma }} \left\{ {f({{\bm{\gamma }}^k}) + \nabla f{{({{\bm{\gamma }}^k})}^T}({\bm{\gamma }} - {{\bm{\gamma }}^k}) + \frac{1}{{2L}}||{\bm{\gamma }} - {{\bm{\gamma }}^k}||_2^2} \right\}\\
		&+ \lambda \sum\limits_{i = 1}^n {\left( {G(|\gamma _i^k|) + g(|\gamma _i^k|)(|\gamma _i^{}| - |\gamma _i^k|)} \right)} \\
		&= \mathop {\arg \min }\limits_{\bm{\gamma }} \left\{ {\frac{1}{{2L}}||{\bm{\gamma }} - ({{\bm{\gamma }}^k} - L\nabla f({{\bm{\gamma }}^k}))||_2^2} \right\} \\
		&+ \lambda \sum\limits_{i = 1}^n {\left( {g(|\gamma _i^k|)(|\gamma _i^{}|)} \right)} \\
		&= \mathop {\arg \min }\limits_{\bm{\gamma }} \left\{ {\frac{1}{{2L}}||{\bm{\gamma }} - ({{\bm{\gamma }}^k} - L{{\bf{A}}^T}({\bf{A}}{{\bm{\gamma }}^k} - {\bf{y}})||_2^2} \right\}\\
		 &+ \lambda \tau \sum\limits_{i = 1}^n {\left( {\frac{{(|\gamma _i^{}|)}}{{|\gamma _i^k| + \tau }}} \right)} 
		\end{aligned}
	\end{equation}
	
	where we remove all irrelevant terms.
	
	(\ref{tuidao_gamma}) is equivalent to ISTA with adaptive thresholds.
	
	\begin{equation}
	\label{ISTA-AT}
		\begin{aligned}
		\theta _i^{k + 1} &= \frac{\varsigma }{{|\gamma _i^k| + \tau }}\\
		\gamma _i^{k + 1} &= {\eta _{\theta _i^{k + 1}}}(\gamma _i^k - L{\bf{A}}_{:,i}^T({\bf{A}}{{\bm{\gamma }}^k} - {\bf{y}}))
		\end{aligned}
	\end{equation}
	
	Where $\varsigma  = \lambda \tau $. The parameters $L$ and $\lambda$ in (\ref{ISTA-AT}) needs to be adjusted manually. For maximizing efficiency and better recovering results, we introduce learnable parameters to enable automatic hyper-parameters tuning as (\ref{LISTA-AT}) demonstrates.
	
	\begin{equation}
	\label{LISTA-AT}
		\begin{aligned}
		\theta _i^{k + 1} &= {\alpha ^{k + 1}}/(|\gamma _i^k| + \tau )\\
		\gamma _i^{k + 1} &= {\eta _{\theta _i^{k + 1}}}(\gamma _i^k - {({\bf{W}}_{:,i}^k)^T}{{\bf{D}}^k})\\
		{{\bf{D}}^{k + 1}} &= {\bf{A}}{{\bm{\gamma }}^{k + 1}} - {\bf{y}}
		\end{aligned}
	\end{equation}
	
	where $\Theta  = \{ {{\bf{W}}^k},{\alpha ^k}\} $ are learnable parameters. And (\ref{LISTA-AT}) is equivalent to LISTA with adaptive element-wise threshold. Same derivation from LISTA to ALSITA, according to~\cite{ALISTA}, we can also apply our element-wise adaptive thresholding method to ALISTA. 
	
	\subsection{ATASI-Net}
	
	The proposed ATASI-Net, as enhanced version of unfolded ISTA based architecture, consists of several well-designed update layers in a cascaded form to pursue a performance boost. Fig. 4 shows the overall architecture of ATASI-Net and illustrates the data flow of each module in the \emph{k}-th layer. As shown, each updating layer consists of one module and three processing units, including the pre-calculation module ($\bf{W}$), the error propagation unit ($\bf{D}$), the threshold updating unit ($\bf{T}$), and the reconstruction unit ($\bf{Z}$). We next introduce the ATASI-Net in detail.
	
	\begin{figure*}
		\centering
		\includegraphics[scale=.6]{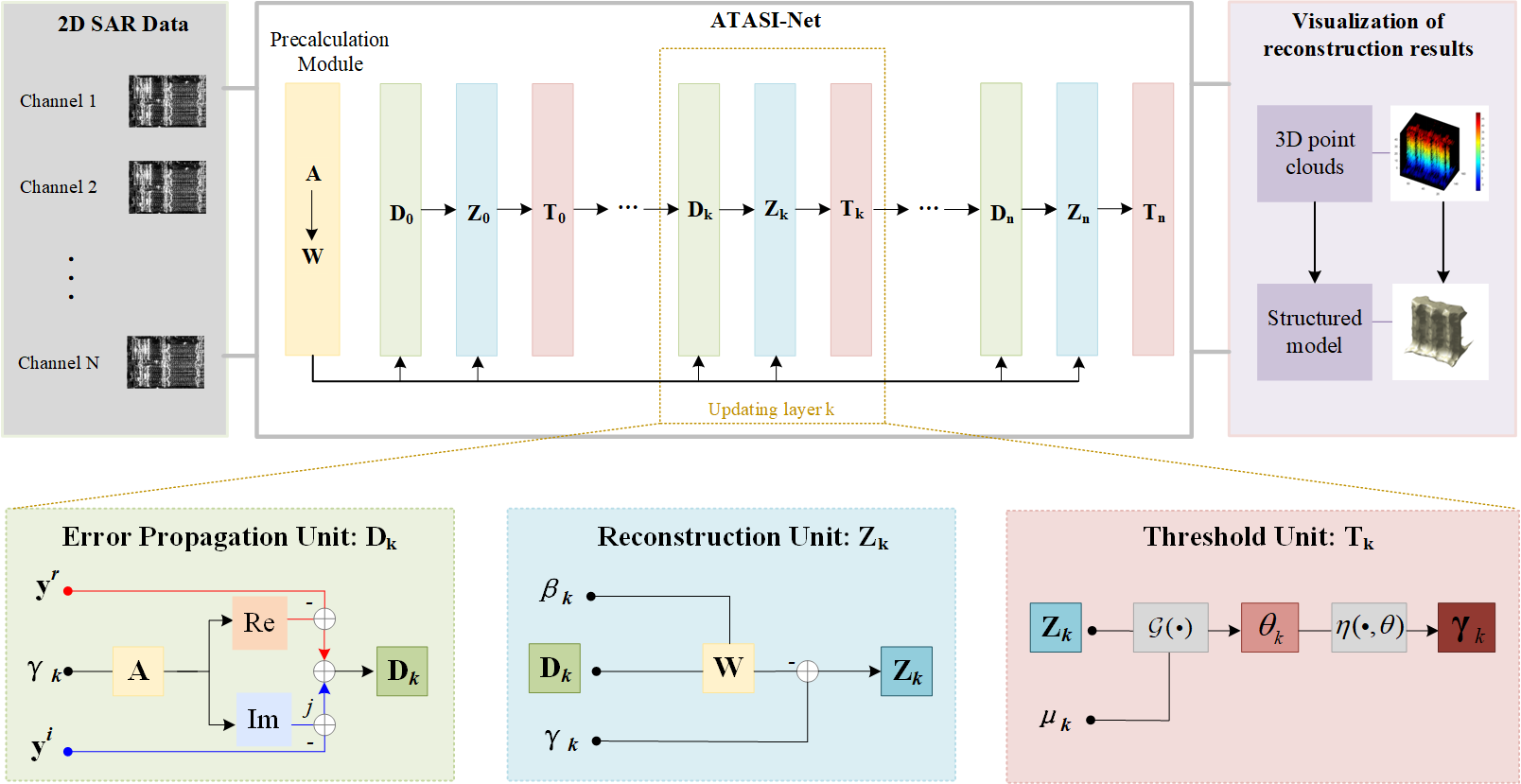}
		\caption{Overall architecture of ATASI-Net and the data flow of each module in the \emph{k}-th layer.}
	\end{figure*}

	1) \emph{Pre-calculation Module} ($\bf{W}$)
	
	Instead of training ${{\bf{W}}^k}$ in (\ref{LISTA-AT}), we can precompute $\mathbf{W}$ off-line by a data-free analytic method, which has been demonstrated in~\cite{ALISTA}. This module computes the fixed weight matric $\bf{W}$ by solving the convex optimization problem:
	
	\begin{equation}
	\label{net-W}
		{\bf{W}} = \mathop {\arg \min }\limits_{{\bf{W}} \in {C^{M \times N}}} ||{{\bf{W}}^T}{\bf{A}}||_F^2{\rm{s}}{\rm{.t}}{\rm{. (}}{{\bf{W}}_{:,i}}{{\rm{)}}^T}{{\bf{A}}_{:,i}} = 1,i = 1,...,n
	\end{equation}
	Where, $\mathbf{A}$ represent the sensing matrix calculated by (3).
	Then, the matrix weight of each layer can be determined just by learning the step size $\aleph$.The pre-calculation module returns the weight matrix ${{\bf{W}}^k}$, for the kth layer:
	
	\begin{equation}
		{{\bf{W}}^k} = {\beta ^k}{\bf{W}}
	\end{equation}
	
	2) \emph{Error Propagation Unit} ($\bf{D}$)
	
	This unit estimates the residual measurement error. As for the output, given the estimated reflection profile of the previous iteration, ${\bf{D}}^k$ can be calculated by
	
	\begin{equation}
		{{\bf{D}}^k} = {\bf{A}}{{\bm{\gamma }}^k} - {\bf{y}}
	\end{equation}
	
	3) \emph{Threshold Unit} ($\bf{T}$)
	
	Different from ALISTA, the threshold of the proposed ATASI-Net is not only layer-varied, but also element-wise, which improved retention against weak targets and utilized the semantic information. Specially, we introduce trainable parameter ${\mu ^k}$ to adjust the step size of the thresholds between adjacent layers. Given the reconstruction signal $z_i^k$ for the \emph{i}-th element, The element-wise shrinkage thresholding operator $\theta_i^k$ can be calculated by (\ref{theta}), and according to [39] the $\varepsilon$ is set to 0.01 to achieve an optimal performance.
	
	\begin{equation}
	\label{theta}
		\theta _i^k = {\mu ^k}\frac{1}{{|z_i^k| + \varepsilon }}
	\end{equation}
	
	The threshold unit ${{\bf{T}}^k}$ computes the shrinkage threshold $\mathbf{\theta}^k$ of the block based on (\ref{theta}).
	
	4) \emph{Reconstruction Unit} ($\bf{Z}$)
	
	The roughly estimated signal can be calculated by a form of one-step gradient descent as:
	
	\begin{equation}
		{{\bf{z}}^{k+1}} = {{\bm{\gamma }}^k} - {\bf{W}}_{:,i}^k{{\bf{D}}^k} = {{\bm{\gamma }}^k} - {\beta ^k}{{\bf{W}}_{:,i}}{{\bf{D}}^k}
	\end{equation}
	
	Then, the reconstruction unit ${\bf{Z}}_{}^k$ obtains the reconstructed reflection profile ${{\bm{\gamma }}^{k + 1}}$ by means of an activation layer defined based on an adaptive threshold shrinkage function, which can be calculated by (\ref{gamma-k})
	
	\begin{equation}
	\label{gamma-k}
		{{\bm{\gamma }}^{k + 1}} = {\eta _{{{\bf{\theta }}^{k + 1}}}}({{\bf{z}}^{k + 1}})
	\end{equation}
	
	For element-wise adaptive thresholds, the specific value of reflection profile $\gamma _i^k$ in the \emph{i}-th amuzith-range cell of the \emph{k}-th layer can be expressed as $\gamma _i^k = {\eta _{\theta _i^k}}(z_i^k) = sign(z_i^k)\max (|z_i^k| - \theta _i^k,0)$.
	
	\setParDis
	5) \emph{Training Strategy and Backpropagation}
	
	Similar to the traditional deep networks, ATASI-Net can be trained from end-to-end via backpropagation in a data-driven manner.
	
	(1) \emph{Loss Function}
	
	To pursuit a high reconstruction quality, mean squared error (MSE) is used as the cost function to obtain optimal hyperparameters. Given dataset used for training ${\Omega _{{\rm{train}}}} = \{ {\bm{\gamma }}_{{\rm{label}}}^k,{\bf{y}}_{}^k\} _{k = 1}^N$ and learnable parameter set ${\bm{\Theta} _{para}} = \{ {\mu _t},{\beta _t}\} _{t = 1}^M$, the cost function can be defined as:
	
	\begin{equation}
		{\mathcal{L}} {\rm{(}}\bm{\Theta} {\rm{)}} = \frac{1}{{{N_{{\rm{train}}}}}}\sum\limits_{{\Omega _{{\rm{train}}}}} {{\rm{||}}\bm{\hat \gamma} (\bm{\Theta} ,{\bf{y}})}  - {\bm{\gamma} _{{\rm{label}}}}{\rm{|}}{{\rm{|}}_{\rm{2}}}
	\end{equation}
	
	Where $N_{train}$ represents the number of trained samples, and ${\bm{\hat \gamma }}(\bm{\Theta} ,{\bf{y}})$ denotes the network output based on the echo signal $\bf{y}$ and network parameters $\bm{\Theta} $.
	
	(2) \emph{Backpropagation and Gradient Calculation}
	
	Similar to typical deep networks, the parameters of ATASI-Net can be optimized via backpropagation with a variety of optimizers, such as stochastic gradient descent (SGD)~\cite{SGD}, Adam~\cite{Adam} and Adadelta~\cite{Adadelta}.
	
	In the training process of the model, the observation signal $\bf{y}$, the measurement matrix $\bf{W}$, and the reflection profile $\bm{\gamma}$ are all complex-valued. Therefore, the proposed ATASI-Net is a complex-valued network, even if the trained parameters are real. Inspired by the work in~\cite{Li2}, the gradients of ${\mathcal{L}} {\rm{(}}\Theta {\rm{)}}$ for the parameters $\Theta $ of the \emph{k}-th layer can be computed via the formulas below:
	
	\begin{equation}
		\frac{{\partial {\mathcal{L}} }}{{\partial \bm{\Theta} _k^T}} = \frac{{\partial {\mathcal{L}} }}{{\partial {\bf{O}}_k^T}}\frac{{\partial {{\bf{O}}_k}}}{{\partial \bm{\Theta} _k^T}}
	\end{equation}
	
	And for the partial derivation of a complex-valued matrix $\mathbf{O}$ can be calculated as follows~\cite{complex}:
	
	\begin{equation}
		\frac{{\partial {\mathcal{L}} }}{{\partial {\bf{O}}}} = \frac{{\partial {\mathcal{L}} }}{{\partial {\rm{Re\{ O\} }}}} + j\frac{{\partial {\mathcal{L}} }}{{\partial {\rm{Im\{ O\} }}}}
	\end{equation}
	
	It should be pointed out that, in widely used deep learning toolboxes such as Pytorch and TensorFlow, the complex numbers are already supported, making it convenient for implementing complex number networks. In this article, the proposed network ATASI-Net is implemented by Tensorflow. The basic workflow of the proposed ATASI-Net is shown in Algorithm 1.
	
	\begin{algorithm}[htb]\vspace{0.2em} 
		\caption{Summary of the proposed ATASI-Net. }
		\label{alg:Framwork}
		\begin{algorithmic}
			\STATE \textbf{1) Obtain training data};\\
			\STATE \quad\textbf{Generate} reflectivity profile $\boldsymbol{\gamma}$\\
			\STATE \quad\textbf{Generate} measurement matrix $\mathbf{A}$ via (3)\\
			\STATE \quad\textbf{Acquire} simulation signal $\bm{y}$ via (2);\\
			\STATE \quad\textbf{Finish} The generation of training data $\{(\bm{y}_i,\bm{\gamma}_i)\}_{i=1}^{M\times N}$\vspace{0.5em}\\
			
			\STATE \textbf{2) Training of ATASI-Net}
			\STATE \quad\textbf{Precomputing $\mathbf{W}$} by solving\vspace{0.5em}\\
			\STATE \quad\quad $\mathbf{W} = \mathop{\arg \min}\limits_{\mathbf{W}\in \mathbb{C}^{M\times N}}\|\mathbf{W}_T\mathbf{A}\|_F^2\ \ \text{s.t.}(\mathbf{W}_{:.i}\mathbf{A}_{:,i},i=1,...,n)$\vspace{0.5em}\\
			\STATE \quad\textbf{Over given training samples}
			$\{(\bm{y}_i,\bm{\gamma}_i)\}_{i=1}^{M\times N}$\vspace{0.5em}\\
			\STATE \quad\quad ${\min\mathcal{L}} (\bm{\Theta}) = \frac{1}{{{N_{{\rm{train}}}}}}\sum\limits_{{\Omega _{{\rm{train}}}}} {{\rm{||}}\hat \gamma (\bm{\Theta} ,{\bf{y}})}  - {\gamma _{{\rm{label}}}}{\rm{|}}{{\rm{|}}_{\rm{2}}}$\vspace{0.5em}\\
			\STATE\quad\quad 
			\STATE \quad\quad where ${\bm{\Theta}} = \{ {\mu _k},{\beta _k}\} _{k = 1}^K$\\\vspace{0.5em}
			\STATE \quad\textbf{Finish} obtain the optimal parameters
			${\bm{\Theta}} = \{ {\mu _k},{\beta _k}\} _{k = 1}^K$\vspace{0.5em}\\
			\STATE \textbf{3) ATASI-Net for TomoSAR}
			\STATE \quad\textbf{Input:} observed signal $\bm{y}$, measurement matrix $\mathbf{A}$
			\STATE \quad\textbf{Output:} reflectivity profile $\bm{\gamma}$, threshold $\bm{\theta}$
			\STATE \quad 1: Load the optimal parameter set ${\bm{\Theta}} = \{ {\mu _k},{\beta _k}\} _{k = 1}^K$,\\
			\quad\quad K is the maximum number of layers;
			\STATE \quad 2: Pre-compute $\mathbf{W}$ via (19)
			\STATE \quad 3: Set $\varepsilon$ = 0.005$\times$$y_{max}$, $\beta^0$ = 0.01; Initialize $\bm{\gamma^0}$, $\bm{D}^0$
			\STATE \quad 4: \textbf{while} k $ \le $ K \textbf{do}
			\STATE \quad 5: \quad ${{\bf{W}}^k} = {\beta ^k}{\bf{W}}$
			\STATE \quad 6: \quad ${{\bf{z}}^{k}} = {{\bm{\gamma }}^{k-1}} - {\bf{W}}_{:,i}^{k-1}{{\bf{D}}^{k-1}}$
			\STATE \quad 7: \quad ${{\bf{D}}^k} = {\bf{A}}{{\bm{\gamma }}^k} - {\bf{y}}$
			\STATE \quad 8: \quad $	\theta _i^k = {\mu ^k}\frac{1}{{|z_i^k| + \varepsilon }}$
			\STATE \quad 9: \quad $		{{\bm{\gamma }}^{k + 1}} = {\eta _{{{\bf{\theta }}^{k + 1}}}}({{\bf{z}}^{k + 1}})$
			\STATE \quad 10: \quad k = k + 1
			\STATE \quad 11: \textbf{end while}
		\end{algorithmic}
	\end{algorithm}
	
	\subsection{Discussion}
	
	The proposed ATASI-Net takes advantages of both the theoretical interpretability of conventional model-driven algorithms and the learnability of deep networks. A modified version of unfolded ISTA is taken as the backbone, which aims to pursue a performance boost. The characteristics of ATASI-Net are discussed as follows.
	
	Compared to the traditional CS algorithm such as OMP and ISTA, the proposed network learns parameters that need to be manually adjusted, such as threshold and step size, by deep learning approach. At the same time, an adaptive threshold is proposed, which enables the parameters to be not only layer-varied but also element-wise. Thus, compared with the traditional threshold shrinkage function the proposed module has better retention ability of small and weak targets, as well as unveiling and utilizing the semantic information.
	
	Compared with conventional sparse microwave reconstruction methods, ATASI-Net show superiority in both speed and accuracy. Firstly, ATASI-Net is designed into a feed-forward network structure, which is suitable to be accelerated by GPU. Meanwhile, the iterative nature of conventional algorithms always makes them suffer from high data dependency and low parallelism. Secondly, ATASI-Net learns optimal parameters automatically in an end-to-end manner instead of a huge effort in tuning, which guarantees efficiency and avoids performance decreasing caused by mistuning of parameters. Thirdly, ATASI-Net only learns the threshold and step parameters, both are real-valued. Therefore, compared with deep networks that need to learn matrix parameters, we can significantly reduce the number of training parameters and system complexity (the number of parameters to learn is the same as ALISTA:$\mathcal{O}(K)$), which in turn reduces the training burden and required training dataset size.

 \section{Simualion Experiments}
 
In the experimental part, simulation and real data are used to illustrate the efficiency and superiority of the proposed network. The simulation experiments include scattering point simulation and 3D building model simulation. The settings of these two simulations are different, and the parameters of the simulation experiment of 3D building facade are the same as those of the actual scene.

\subsection{Experiments Settings}

The training samples in our simulation include two types, one is the single scatterer sample with only one target point in the same range-azimuth cell, and the other is the double scatterer sample with overlapping mask in the same range-azimuth cell. 

For a single scatterer, the backscattering coefficient is a complex number, which can be expressed as $\gamma  = A\exp (j\phi )$ where the amplitude $A$ are randomly distributed from 0 to 4 following Rayleigh distribution and the scattering phase $\phi $ are randomly distributed from 0 to 2$\pi$ following uniform distribution. With the provided parameters and the known imaging geometry, the measurement matrix $\bf{D}$ can be determined following (3). The number of discretized grids in each range-azimuth cell along the elevation is fixed to 128. For double scatterers, the generation of the two single scatterers is identical to the previous step. In addition, for double scatterers, we also vary the elevation distance between the two single scatterers from 0.1 to 1.2 times of Rayleigh resolution. The echo signal is simulated at 11 different SNR levels between [0, 30 dB] with additive white Gaussian noise.

\subsection{ATASI-Net Configuration and Training}

All the experiments are conducted on a platform with 4.70GHz Intel Core i7-12700H CPU and NVIDIA RTX 3060 GPU (6G Memory). During the training stage, the ATASI-Net is optimized with Adam optimizer. The learning rate is set to 0.001, which decays dynamically according to the loss descent rate in the training process.

\subsection{Scatterer Simulation Experiment}

The simulation parameters of the scatterer experiment were set as shown in Table I, For uniform linear arrays, the expected Rayleigh elevation resolution ${\rho _s}$ can be calculated by ${\rho _s} = \lambda r/2n\Delta d$,  which equals to 50m in this simulation. The training data set of this scatterer simulation experiment is 20000, half of which are single scatterers and the other half are double scatterers, and the training data was generated as described above.

\begin{table*}[]
	\caption{SYSTEM PARAMETERS OF SIMULATIONS AND REAL EXPERIMENTS}
	\label{parameter}
	\centering
	\renewcommand{\arraystretch}{1.5}
	\begin{tabular}{|c|c|c|c|}
		\hline
		Parameter                         & Scatter Simulation                           & 3D Building Simulation                       & SARMV3D1.0                                          \\ \hline
		Carrier frequency $f_c$           & 14.25 GHz                                    & 14.25 GHz                                    & 14.25 GHz                                           \\ \hline
		Wavelength $\lambda$              & 0.021 m                                      & 0.021 m                                      & 0.021 m                                             \\ \hline
		Antenna interval $\Delta d$       & 0.084 m                                      & 0.084 m                                      & 0.084 m                                             \\ \hline
		Array number $n$                  & 8                                            & 8                                            & 8                                                   \\ \hline
		Range $R_0$                       & 400 m                                        & 1200 m                                       & 1184-1304 m                                         \\ \hline
		Average incident angle $\theta_0$ & 45$^{\circ}$ & 30$^{\circ}$ & 24.9-34.5$^{\circ}$ \\ \hline
	\end{tabular}
\end{table*}

\begin{figure}[!h]
	\centering
	\includegraphics[width=2.5in]{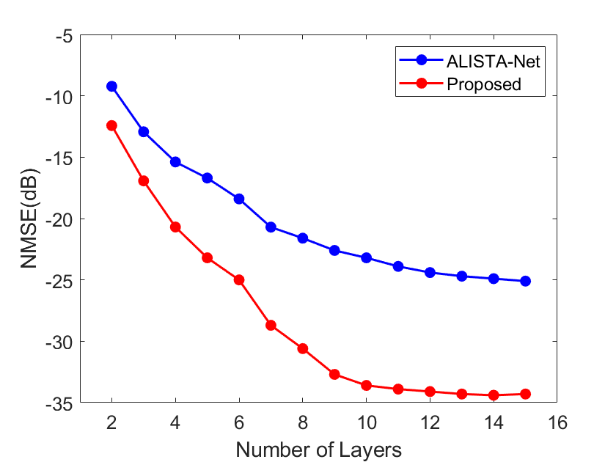}
	\caption{Comparison of performance w.r.t the number of layers between ALISTA and proposed ATASI-Net. The proposed method conduces to faster convergence and improves the estimation accuracy.}
	\label{layer}
\end{figure}

In the training procedure, we gradually increased the number of layers from 2 to 15 to determine an optimal network structure and compared conventional ALISTA with the proposed ATASI-Net in terms of the normalized MSE (NMSE). The validation dataset contains 2000 samples simulated in the presence of noise with SNR = 30dB using the same settings as shown in Table 1. Fig. \ref{layer} illustrates the performance of ALISTA and proposed network. Closer inspection of Fig. \ref{layer}  shows that the proposed ATASI-Net has smaller NMSE, better network performance and faster convergence speed than ALISTA with the same layer number. Meanwhile, it can be seen that when the proposed network has more than 10 layers, the performance payback of layer number will be diminished, as well as the training burden will be increased. Therefore, the 10-layer network is used in our experiment. 

\subsection{Double Scatters Simulation Analysis}

For the double scatterer simulation experiment, we implemented four different algorithms, including SVD, OMP, ALISTA, and proposed ATASI-Net, for comparison. It is worth noting that the four methods represent three types of paradigms, including traditional spectrum estimation, the conventional CS-based algorithm, and the deep-network-based method. By changing the SNR and the distance between the two scatterers, \emph{six groups of 1000 samples each} were used to analyze the simulation experiments of the double scatterers with different SNRs and different spacing. We employed normalized distance $\alpha$ to represent the distance between the two scatterers, which is defined as follows:

\begin{equation}
	\alpha  = \frac{{{d_s}}}{{{\rho _s}}}
\end{equation}

Where $d_s$ is the actual distance between the two scatterers along the elevation direction.

\begin{figure*}[t]
	\centering
	\subfigure[]{\includegraphics[width=0.3\textwidth]{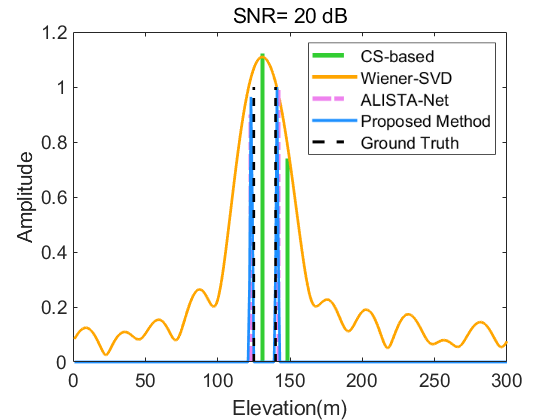}}\vspace*{0em}
	\subfigure[]{\includegraphics[width=0.3\textwidth]{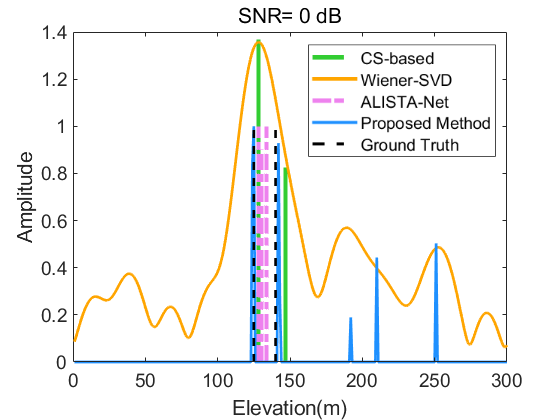}}\vspace*{0em}
	\subfigure[]{\includegraphics[width=0.3\textwidth]{rho0.3_2.png}}\vspace*{0em}
	\subfigure[]{\includegraphics[width=0.3\textwidth]{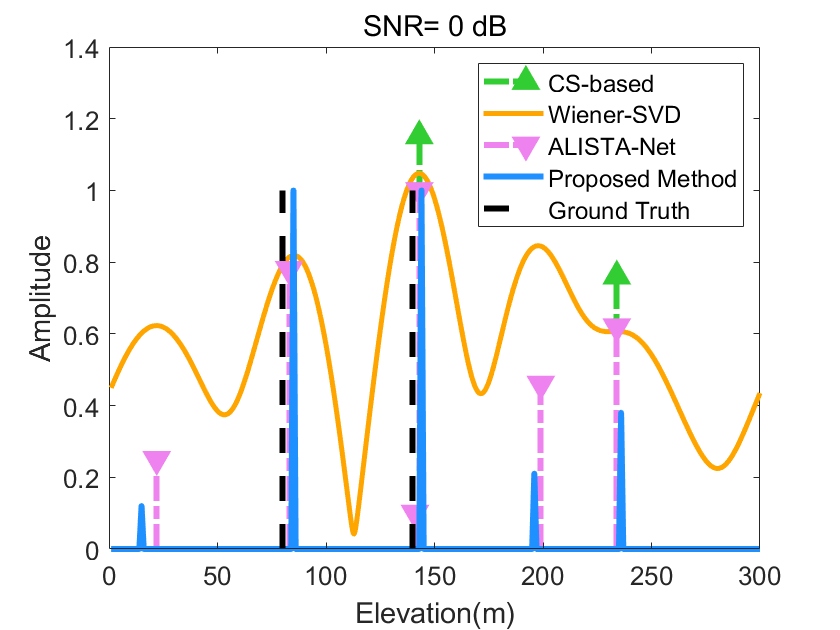}}\vspace*{0em}
	\subfigure[]{\includegraphics[width=0.3\textwidth]{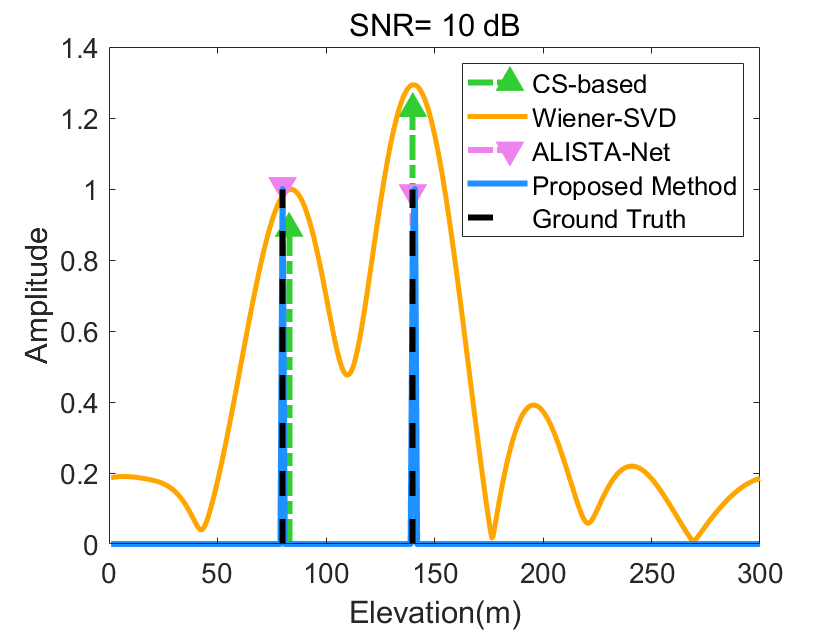}}\vspace*{0em}
	\subfigure[]{\includegraphics[width=0.3\textwidth]{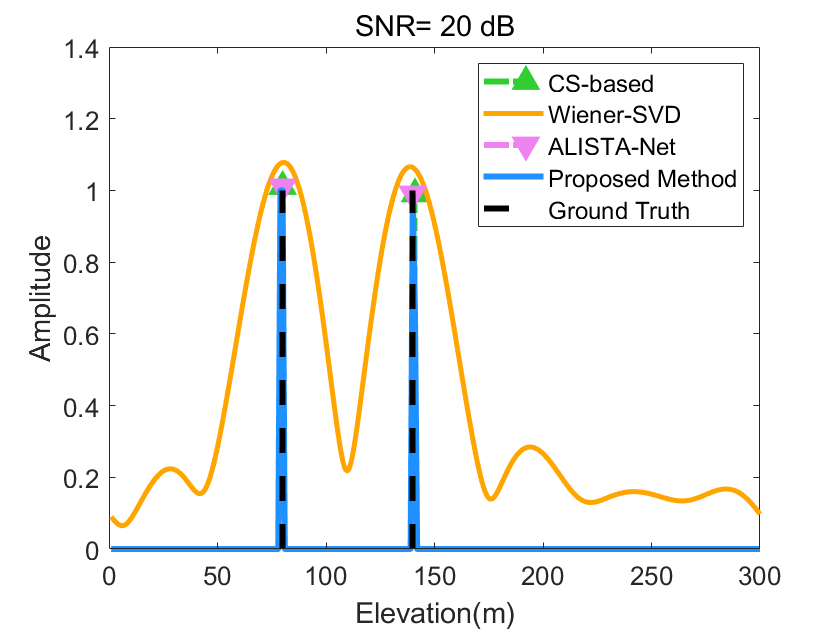}}\vspace*{0em}
	\caption{Estimated reflectivity profile of simulation with double scatterers for a range-azimuth cell. (a) $alpha$ = 0.3$\rho$, SNR = 0 dB (b) $alpha$ = 0.3$\rho$, SNR = 10 dB (c) $alpha$ = 0.3$\rho$, SNR = 20 dB (d) $alpha$ = 1.2$\rho$, SNR = 0 dB (e) $alpha$ = 1.2$\rho$, SNR = 10 dB (f) $alpha$ = 1.2$\rho$, SNR = 20 dB}
	\label{double-scatters}
\end{figure*}

Fig. \ref{double-scatters} demonstrates the reconstruction results of different algorithms under different SNR and $\alpha$.

As we can see from Fig. \ref{double-scatters}, when $\alpha=1.2$, all four algorithms can successfully reconstruct the scatterers. However, when $\alpha=0.3$, the traditional CS-based greedy algorithm, and SVD spectrum estimation method cannot distinguish the targets with close distance, and only the sparse unfolded network ALISTA and the proposed ATASI-Net can distinguish them. In addition, when SNR is high (20dB), all four methods can obtain relatively accurate target information for the two scatterers in the same range-azimuth cell. And when the SNR is reduced to 10 dB, although all four methods can reconstruct the position information of the two scatterers in the elevation, the traditional spectrum estimation and CS-based method have some errors in the estimation of reflection intensity, and the method based on the sparse unfolded deep network can estimate the reflection information more accurately. Particularly, when the SNR = 0 dB, the traditional CS-based greedy algorithm, and the spectrum estimation method of SVD lose their anti-noise ability and cannot estimate the target position, and scattering information. Moreover, the sparse unfolded deep network ALISTA has a certain deviation in the estimation of the target position along the elevation, which is within an acceptable range. The accuracy of the proposed ATASI-Net for the position estimation of the two scatterers is similar to that of ALISTA, but the anti-noise performance of the proposed ATASI-Net is better than the other three methods, because the adaptive threshold unit can filter the noise without affecting the retention of the target.

To more clearly illustrate the performance of the proposed ATASI-Net, we defined the detection success rate according to the literature [33] [42], and the traditional CS-based method was compared with our proposed ATASI-Net for further analysis. The detection success rate should satisfy that the estimated elevation of the two scatterers should be within both $ \pm $ times Cramér–Rao lower bound (CRLB) and $\pm$ 0.5 times $d_s$ for their true elevation. For the latter condition, which needs to be satisfied, it mainly acts as a constraint when the two scatterers are in close proximity.

According to [42], the CRLB ${\sigma _d}$ of the two scatterers can be calculated by:

\begin{equation}
	{\sigma _d} = {c_0} \cdot {\sigma _s}
\end{equation}

Where

\begin{equation}
	{\sigma _s} = \frac{{\lambda {R_0}}}{{4\pi \sqrt {2N \cdot SNR}  \cdot {\sigma _b}}}
\end{equation}

is the CRLB of the elevation estimates of single scatterer. ${\sigma _b}$ is the standard deviation of the elevation aperture sample positions, and for uniform linear array ${\sigma _b} = {\rho _s}/\sqrt {12} $, and by the empirical formula:

\begin{footnotesize}
\begin{equation}
	{c_0} = \sqrt {\max \left\{ {\frac{{40{\alpha ^{ - 2}}(1 - \alpha /3)}}{{9 - 6(3 - 2\alpha )\cos \left( {2\Delta \varphi  + 2\pi \alpha \left( { - \frac{1}{N}} \right)} \right) + {{(3 - 2\alpha )}^2}}},1} \right\}} 
\end{equation}
\end{footnotesize}

Where $\Delta \varphi $ is the phase difference between the two scatterers.

\begin{figure}[h]
	\centering
	\includegraphics[width=2in]{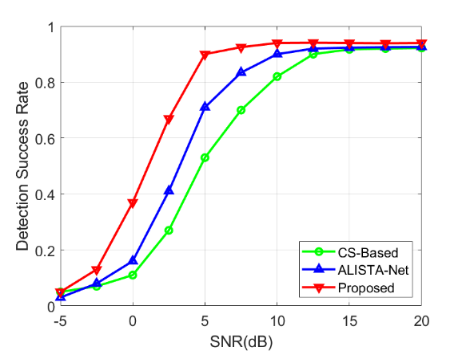}
	\caption{Detection success rate (DSR) as a function of SNR under the case: $\alpha = 0.6$.}
	\label{SNR-DSR}
\end{figure}

\begin{figure}[!h]
	\centering
	\includegraphics[width=2in]{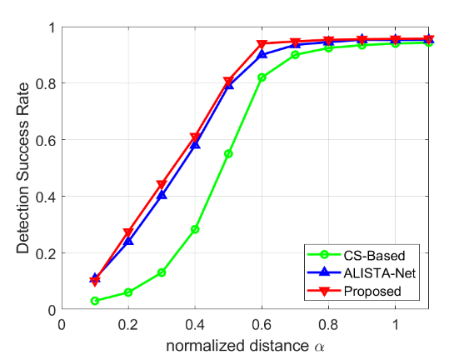}
	\caption{Detection success rate (DSR) as a function of $\alpha$ under the case: SNR = 10 dB.}
	\label{alpha-DSR}
\end{figure}

We compared the detection success rates of different SNR, $\alpha$, amplitude ratio, and phase differences respectively. Moreover, for each experiment, the Monte Carlo experiments with one thousand trials were executed. Firstly, we fixed the normalized distance $\alpha$ with 0.6, gradually increase SNR from -5 dB to 20 dB, and compared the performance of the CS-based method, the sparse unfolded network ALISTA and our proposed ATASI-Net, and the results are shown in Fig. \ref{SNR-DSR}. It can be seen that the proposed ATASI-Net has stronger robustness and the anti-noise performance is better than the other two methods. Secondly, we compared the detection success rates of different algorithms under different $\alpha$ with SNR = 10 dB. As can be seen from Fig. \ref{alpha-DSR}, the method based on the sparse unfolded network is superior to the traditional CS-based algorithm in the reconstruction of position information of the target with a short distance. 

\begin{figure}[!h]
	\centering
	\includegraphics[width=2in]{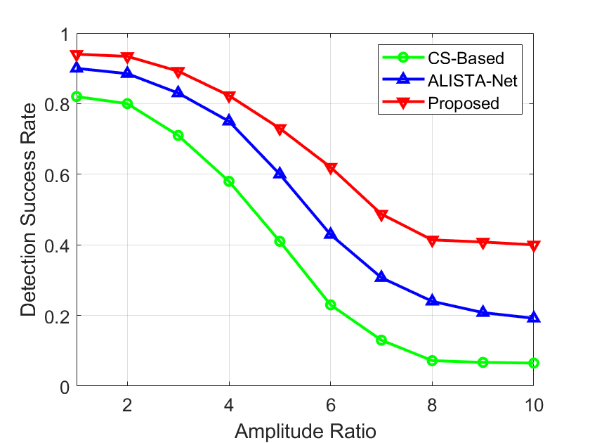}
	\caption{Detection success rate (DSR) as a function of amplitude ratio under the case: $\alpha = 0.6$, SNR = 10 dB.}
	\label{apm-DSR}
\end{figure}

Thirdly, we evaluated the performance of the proposed ATASI-Net at different amplitude ratios of the double scatterers with fixed SNR = 10dB, $\alpha$=0.6. As the amplitude ratio of the two scatterers increases, the scatterer with a smaller amplitude becomes less prominent. Therefore, at high amplitude ratios, the detection success rate will decrease. However, our proposed ATASI-Net has a higher detection success rate for double scatterers with different amplitude ratios as Fig. \ref{apm-DSR}. 

\begin{figure}[!h]
	\centering
	\includegraphics[width=2in]{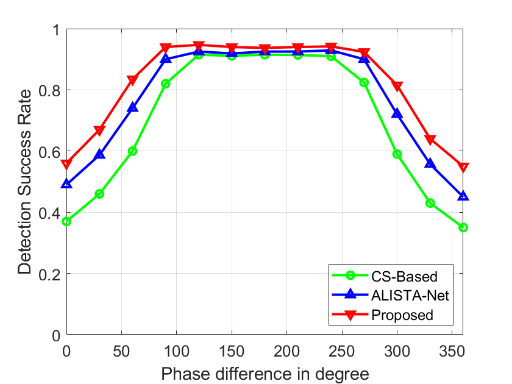}
	\caption{Detection success rate (DSR) as a function of phase difference under the case: $\alpha = 0.6$, SNR = 10 dB.}
	\label{phase-DSR}
\end{figure}

Fourthly, the phase difference between the double scatterers was changed in the simulation experiments to further verify the performance of the proposed ATASI-Net. Fig. \ref{phase-DSR} shows the detection success rate when the SNR = 10 dB and the normalized distance $\alpha$=0.6. The double scatterers in the simulation were set to have the same amplitude. We can see that the detection success rate of the double scatterers is above 50$\%$ under different phase differences.

\subsection{3D Building Simulation}

The parameters of the 3D building simulation experiment were set as shown in Table. \ref{parameter}. These parameters were approximately the same as the parameters of the real data. 

\begin{figure}[!h]
	\centering
	\includegraphics[width=0.25\textwidth,height=0.16\textheight]{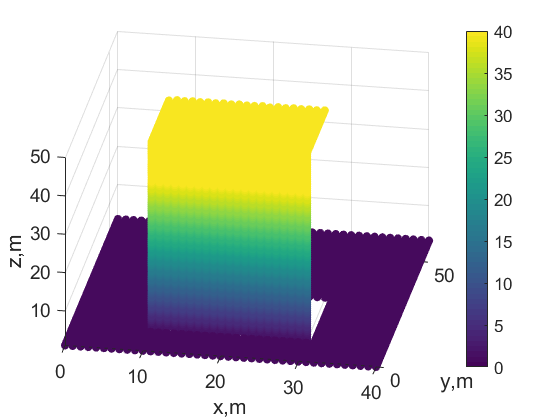}
	\caption{The 3D building scatterer model.}
	\label{3Dmodel}
\end{figure}

In this experiment, a 3D scatterer model of a building with the roof, wall, and ground was constructed to visually compare the imaging performance of different algorithms. Fig. \ref{3Dmodel} displays the 3D views of the simulated target. We compared the OMP algorithm, ISTA algorithm, ALISTA, and the proposed ATASI-Net. It is of note that the first three methods represent three different algorithms: the greedy-based traditional CS algorithm, the $L_1$ regularization-based traditional CS algorithm, and the sparse unfolded deep network, respectively. By changing the SNR, roof: wall: ground RCS ratios and the number of targets (NoTs), we compared the different methods regarding their qualitative visual effect and quantitative evaluation indicator. The specific experimental details are as follows.

1) \emph{Visual comparisons}

\begin{figure*}[t]
	\centering
	\subfigure[]{\includegraphics[width=0.23\textwidth]{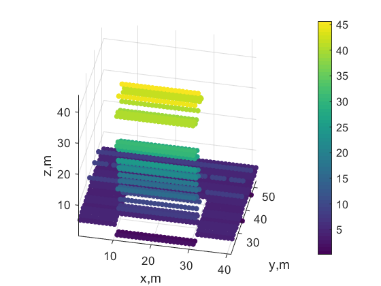}}\vspace*{0em}
	\subfigure[]{\includegraphics[width=0.23\textwidth]{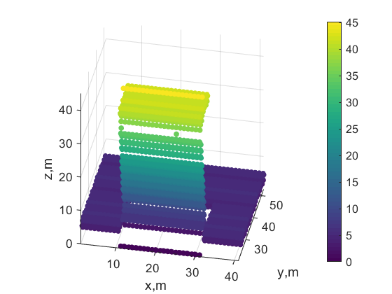}}\vspace*{0em}
	\subfigure[]{\includegraphics[width=0.23\textwidth]{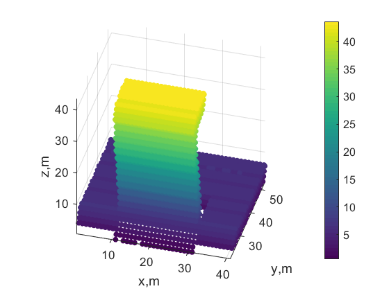}}\vspace*{0em}
	\subfigure[]{\includegraphics[width=0.23\textwidth]{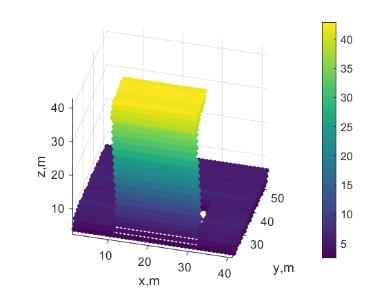}}\vspace*{0em}
	\subfigure[]{\includegraphics[width=0.23\textwidth]{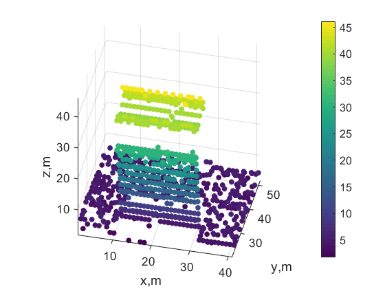}}\vspace*{0em}
	\subfigure[]{\includegraphics[width=0.23\textwidth]{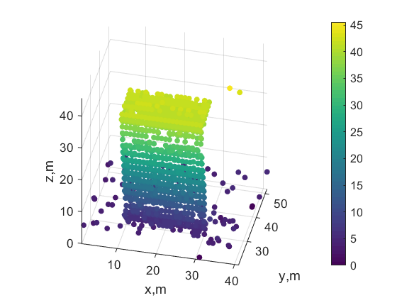}}\vspace*{0em}
	\subfigure[]{\includegraphics[width=0.23\textwidth]{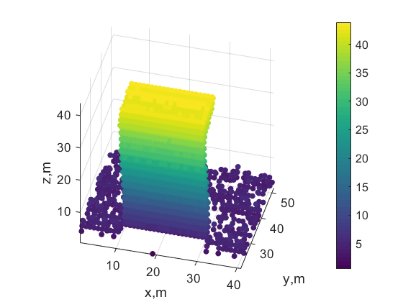}}\vspace*{0em}
	\subfigure[]{\includegraphics[width=0.23\textwidth]{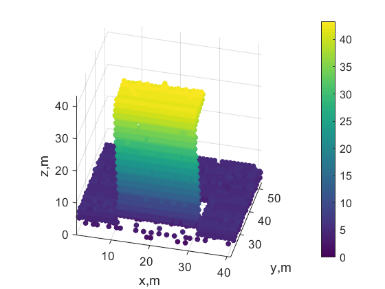}}\vspace*{0em}
	\caption{Imaging results in different noise conditions. Rows 1-2 correspond to SNR = 30 dB and SNR = 10 dB. Columns 1–4 correspond to the results of OMP, ISTA, ALISTA and the proposed method, respectively.}
	\label{SNR-3D}
\end{figure*}

To compare various sparse reconstruction algorithms in terms of imaging quality, the roof: wall: ground RCS ratio was fixed to 2: 2: 1 and NoT was fixed to 7000, while the SNRs were different. The results obtained are displayed in Fig. \ref{SNR-3D}. Noticeably, both ALISTA and the proposed ATASI-Net were capable of reconstructing the 3D target precisely when SNR = 30 dB. Nevertheless, the reconstruction quality decreased as SNR decreased. Deep network-based methods appeared to yield images with intact profiles compared with conventional sparse reconstruction algorithms, like OMP and ISTA. This is expected because network-based methods automatically learn the optimal parameters instead of requiring manual tuning, which avoids the performance deterioration caused by parameter mistuning. However, when the SNR decreased to 10 dB, OMP, ISTA, as well as ALISTA yielded images with noisy backgrounds. By contrast, the proposed ATASI-Net was still able to produce a well-reconstructed image, thanks to its element-wise threshold.

\begin{figure*}[hptb]
	\centering
	\subfigure[]{\includegraphics[width=0.23\textwidth]{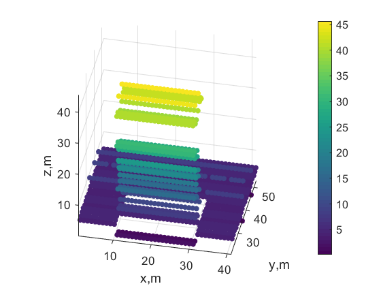}}\vspace*{0em}
	\subfigure[]{\includegraphics[width=0.23\textwidth]{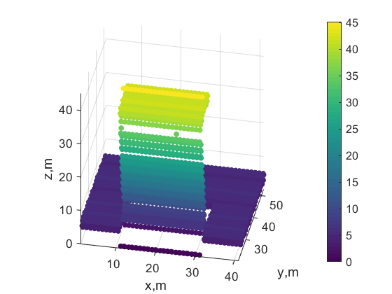}}\vspace*{0em}
	\subfigure[]{\includegraphics[width=0.23\textwidth]{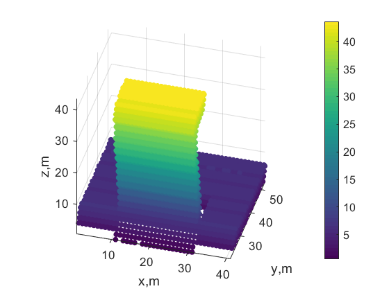}}\vspace*{0em}
	\subfigure[]{\includegraphics[width=0.23\textwidth]{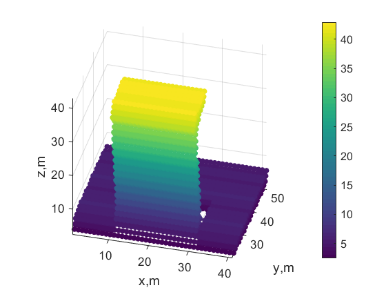}}\vspace*{0em}
	\subfigure[]{\includegraphics[width=0.23\textwidth]{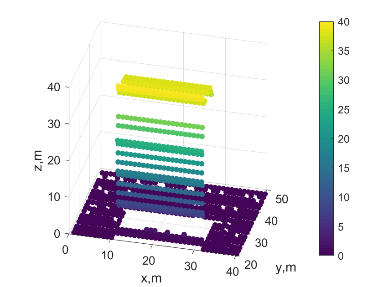}}\vspace*{0em}
	\subfigure[]{\includegraphics[width=0.23\textwidth]{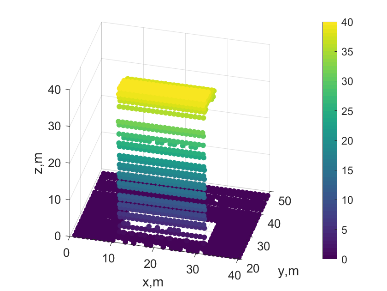}}\vspace*{0em}
	\subfigure[]{\includegraphics[width=0.23\textwidth]{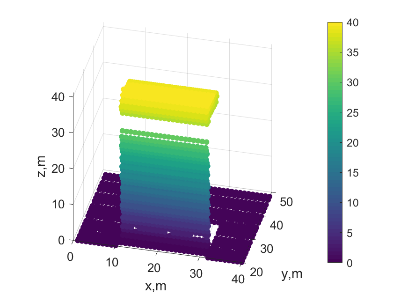}}\vspace*{0em}
	\subfigure[]{\includegraphics[width=0.23\textwidth]{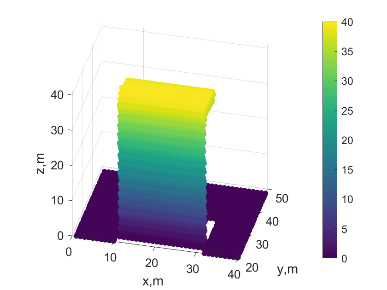}}\vspace*{0em}
	\caption{Imaging results with different roof: wall : ground RCS ratios. Rows 1-2 correspond to roof: wall : ground RCS ratio = 2: 2: 1 and roof: wall : ground RCS ratio = 4: 2: 1. Columns 1–4 correspond to the results of OMP, ISTA, ALISTA and the proposed method, respectively.}
	\label{RCS-3D}
\end{figure*}

Hence, the SNR was fixed to 30 dB, and the images obtained with the different algorithms were compared by applying various roof: wall: ground RCS ratios. As a result, when the RCS ratios differed greatly, the outcome of the reconstruction worsened. In the same resolving unit, scatterer targets with smaller RCS were ignored. Among the four algorithms, the proposed ATASI-Net produced the most compelling imagery results in all cases, demonstrating its superiority. The visual results displayed in Fig. \ref{RCS-3D} are also consistent with the numerical evaluations in the scatterer simulation experiment, which also proves the robustness of ATASI-Net.

2) \emph{Numerical analysis}

In addition to a visual assessment, we also compared the performance of the different algorithms through numerical analysis . The following evaluation metrics were used: peak signal-to-noise ratio (PSNR), normalized averaged root mean square error (NRMSE), and structural similarity index for measuring (SSIM). The first two evaluation indices  measure the accuracy of the reconstructed scattering intensity, and the last evaluation index measures the overall reconstruction effect resulting from the spatial similarity. The different evaluation indicators are defined as follows:

The definition of PSNR is:

\begin{equation}
	\begin{aligned}
	{\rm{MSE}} &= \frac{{\left\| {{\bf{\hat X}} - {{\bf{X}}_{{\rm{ref}}}}} \right\|_F^2}}{{{\rm{num}}\left\{ {{{\bf{X}}_{ref}}} \right\}}}\\
	{\rm{PSNR}} &=  - 10{\log _{10}}\left( {{\rm{MSE}}\left\{ {{\bf{\hat X}},{{\bf{X}}_{ref}}} \right\}} \right)
	\end{aligned}
\end{equation}

where ${\bf{\hat X}}$ is the image reconstructed by the network, ${{\bf{X}}_{ref}}$ is the label image, and num{${{\bf{X}}_{ref}}$} indicates the total pixel number in the image ${{\bf{X}}_{ref}}$. In addition, the definition of SSIM is:

\begin{equation}
	\begin{aligned}
	{\rm{SSIM}} &= \frac{{\left( {2{\mu _{\bf{x}}}{\mu _{\bf{y}}} + {c_1}} \right)\left( {2{\sigma _{{\bf{xy}}}} + {c_2}} \right)}}{{\left( {\mu _{\bf{x}}^2 + \mu _{\bf{y}}^2 + {c_1}} \right)\left( {\sigma _{\bf{x}}^2 + \sigma _{\bf{y}}^2 + {c_2}} \right)}}\\
	{c_1} &= {\left( {{k_1}L} \right)^2},\quad {c_2} = {\left( {{k_2}L} \right)^2}
	\end{aligned}
\end{equation}

where $x,y,{\mu _x},{\mu _y}$ are the reconstructed image, reference image, and their mean values, respectively. ${\sigma _x},{\sigma _y}$ are the standard deviations of $x$ and $y$, respectively, ${\sigma _{xy}}$ represents their cross-covariance. 

Moreover, $L$ stands for the dynamic range of the pixel values. By default, $k_1$ and $k_2$ are set to 0.01 and 0.03, respectively. SSIM is set between 0 and 1. The larger SSIM is, the smaller the gap between the reconstructed image and the reference image, that is, the better the reconstruction quality.

Here, the RCS ratio and NoTs were fixed at 2: 2: 1 and 7000, respectively. The robustness of the different methods was evaluated with SNR values ranging from 0 to 30 dB, with incremental steps of 10 dB. The comparison results are shown in Table. \ref{pingjia-SNR}, and the best evaluations are marked in boldface. The indices show that ATASI-Net achieved the best scores in most conditions, validating the robustness of the proposed method.

\begin{table*}[!hptb]
	\caption{Comparisons of PSNR (dB), NRMSE and SSIM with different SNRs(dB)}
	\label{pingjia-SNR}
	\centering
	\renewcommand{\arraystretch}{1.7}
	\begin{tabular}{lcccccccccccc}
		
		\hline
		\multirow{2}{*}{\textbf{SNR}} & \multicolumn{3}{c}{\text{OMP}}               & \multicolumn{3}{c}{\text{IST}}               & \multicolumn{3}{c}{\text{ALISTA}}             & \multicolumn{3}{c}{\text{Proposed}}             \\
		& \textbf{PSNR} & \textbf{NRMSE} & \textbf{SSIM} & \textbf{PSNR} & \textbf{NRMSE} & \textbf{SSIM} & \textbf{PSNR} & \textbf{NRMSE} & \textbf{SSIM}  & \textbf{PSNR}   & \textbf{NRMSE} & \textbf{SSIM}  \\		\hline
		30                            & 25.739        & 0.261          & 0.909         & 32.429        & 0.117          & 0.946         & 37.283        & 0.076          & \textbf{0.991} & \textbf{40.148} & \textbf{0.061} & 0.990          \\
		20                            & 24.251        & 0.344          & 0.867         & 32.214        & 0.227          & 0.855         & 37.034        & 0.083          & 0.910          & \textbf{40.106} & \textbf{0.056} & \textbf{0.959} \\
		10                            & 22.483        & 0.676          & 0.731         & 31.586        & 0.468          & 0.744         & 36.257        & 0.272          & 0.840          & \textbf{40.074} & \textbf{0.203} & \textbf{0.904} \\
		0                             & 21.179        & 1.063          & 0.522         & 30.773        & 1.079          & 0.524         & 35.068        & 0.664          & 0.647          & \textbf{40.065} & \textbf{0.529} & \textbf{0.768} \\ 		
		Avg.                          & 23.413        & 0.586          & 0.757         & 31.751        & 0.473          & 0.767         & 36.411        & 0.274          & 0.847          & \textbf{40.099} & \textbf{0.212} & \textbf{0.905} 		\\ \hline
	\end{tabular}
\end{table*}

Table. \ref{pingjia-RCS} displays the reconstruction results obtained with the different algorithms, applying different roof: wall: ground RCS ratios, with SNR being 30 dB and NoTs being fixed to 7000.  As expected, for the other three methods with traditional fixed thresholds, as the RCS ratio increased, the difference in scattering intensity between the target points  increased, and thus the indices deteriorated. By contrast, in our proposed ATASI-Net, the thresholds showed an element-wise adaptation, so that the weak target points were not directly removed by threshold shrinkage, thus improving the retention of weak target points to some extent.

\begin{table*}[!hptb]
	\caption{Comparisons of PSNR (dB), NRMSE and SSIM with different roof: wall: ground RCS ratios}
	\label{pingjia-RCS}
	\centering
	\renewcommand{\arraystretch}{1.7}
	\begin{tabular}{ccccccccccccc}
		\hline
		\multicolumn{1}{l}{\multirow{2}{*}{\textbf{RCS}}} & \multicolumn{3}{c}{OMP}                        & \multicolumn{3}{c}{IST}                        & \multicolumn{3}{c}{ALISTA}                      & \multicolumn{3}{c}{Proposed}                      \\
		\multicolumn{1}{l}{}                              & \textbf{PSNR} & \textbf{NRMSE} & \textbf{SSIM} & \textbf{PSNR} & \textbf{NRMSE} & \textbf{SSIM} & \textbf{PSNR} & \textbf{NRMSE} & \textbf{SSIM}  & \textbf{PSNR}   & \textbf{NRMSE} & \textbf{SSIM}  \\ \hline
		1:1:1                                             & 27.173        & 0.227          & 0.922         & 33.852        & 0.101          & 0.957         & 38.624        & 0.062          & \textbf{0.996} & \textbf{40.739} & \textbf{0.063} & 0.994          \\
		2:2:1                                             & 25.739        & 0.261          & 0.909         & 32.429        & 0.117          & 0.946         & 37.283        & 0.076          & 0.991          & \textbf{40.148} & \textbf{0.061} & \textbf{0.990} \\
		4:2:1                                             & 20.479        & 0.334          & 0.827         & 27.633        & 0.153          & 0.803         & 31.646        & 0.129          & 0.836          & \textbf{37.518} & \textbf{0.082} & \textbf{0.902} \\
		9:3:1                                             & 11.127        & 0.347          & 0.512         & 17.941        & 0.186          & 0.469         & 20.559        & 0.155          & 0.505          & \textbf{30.294} & \textbf{0.106} & \textbf{0.747} \\
		Avg.                                              & 21.130        & 0.292          & 0.793         & 27.964        & 0.139          & 0.794         & 32.028        & 0.106          & 0.832          & \textbf{37.175} & \textbf{0.078} & \textbf{0.908} \\ \hline
	\end{tabular}
\end{table*}

Afterward, we experimentally compared the performance of the four kinds of algorithms when applying different NoTs. The RCS ratio and SNR were set to 2: 2: 1 and 30 dB, respectively. The NoTs were changed to the range of 1000–9000 with an incremental step of 2000. A comparison of the PSNR, NRMSE, SSIM and Runtime scores is shown in Table. \ref{pingjia-NoT}. 

The results of the runtime comparison show that as the number of target points increased, the runtime of the traditional CS-based algorithm also increased. Thus, the time-space complexity of the solution  was high for the actual large-scale scenario. For the network-based methods, once the pre-training was completed, little computational time was required compared with conventional CS-driven algorithms. This may be explained by the pre-training process shifting the computational burden to the learning stage of the parameters, consequently yielding a high reconstruction efficiency.

\begin{table*}[]
	\caption{Comparisons of PSNR (dB), NRMSE, SSIM and Runtime (s) with different different number of targets (NoTs)}
	\label{pingjia-NoT}
	\centering
	\renewcommand{\arraystretch}{1.7}
	\resizebox{\textwidth}{!}{
	\begin{tabular}{ccccccccccccccccc}
		\hline
		\multirow{2}{*}{\textbf{NoTs}} & \multicolumn{4}{c}{OMP}                                           & \multicolumn{4}{c}{IST}                                           & \multicolumn{4}{c}{ALISTA}                                         & \multicolumn{4}{c}{Proposed}                                         \\
		& \textbf{PSNR} & \textbf{NRMSE} & \textbf{SSIM} & \textbf{Runtime} & \textbf{PSNR} & \textbf{NRMSE} & \textbf{SSIM} & \textbf{Runtime} & \textbf{PSNR} & \textbf{NRMSE} & \textbf{SSIM}  & \textbf{Runtime} & \textbf{PSNR}   & \textbf{NRMSE} & \textbf{SSIM}  & \textbf{Runtime} \\
		\hline
1000 & 35.428 & 0.217 & 0.943  & \textbf{0.121} & 39.937  & 0.093  & 0.963 & 0.526  & 42.339  & 0.048  & 0.994          & 0.223           & \textbf{43.837}  & \textbf{0.046}  & \textbf{0.994} & 0.242  \\
3000 & 33.774 & 0.221 & 0.936  & \textbf{0.212} & 38.241  & 0.098  & 0.967 & 0.716  & 41.173  & 0.054  & 0.992          & 0.267           & \textbf{42.559}  & \textbf{0.051}  & \textbf{0.992} & 0.283  \\
5000 & 29.636 & 0.239 & 0.927  & 0.397          & 36.456  & 0.104  & 0.955 & 0.924  & 39.836  & 0.062  & \textbf{0.990} & \textbf{0.219}  & \textbf{40.736}  & \textbf{0.057}  & 0.989          & 0.279  \\
7000 & 25.739 & 0.261 & 0.909  & 0.584          & 32.429  & 0.117  & 0.946 & 1.109  & 37.283  & 0.076  & \textbf{0.991} & \textbf{0.245}  & \textbf{38.148}  & \textbf{0.061}  & 0.990          & 0.316  \\
9000 & 16.238 & 0.307 & 0.873  & 0.894          & 27.533  & 0.152  & 0.904 & 1.644  & 33.492  & 0.091  & \textbf{0.933} & \textbf{0.242}  & \textbf{35.169}  & \textbf{0.078}  & 0.924          & 0.289  \\
Avg. & 28.163 & 0.249 & 0.918 & 0.442         & 34.920 & 0.113 & 0.947 & 0.984 & 38.825 & 0.066 & \textbf{0.980}  & \textbf{0.239} & \textbf{40.090} & \textbf{0.059} & 0.978         & 0.282 \\ \hline
	\end{tabular}}
\end{table*}

\section{Real Data Experiment}

In this section, we analyze the applicability of the proposed ATASI-Net for TomoSAR imaging  with real radar data. The visual reconstructions based on both the traditional CS algorithm and the proposed method were compared, and the 3D reconstruction results were expressed as point clouds. At the same time, structured modeling of the 3D point clouds was carried out to achieve a thorough comparison. In addition, we sliced the reconstruction results of the ground  obtained with both the traditional CS algorithm and ATASI-Net and compared them. Moreover, we used the portion that exhibited the greatest focusing effect in the reconstruction from the traditional CS algorithm  as the training set. This portion corresponded to points in the last building, and the aim was to evaluate the reconstruction results obtained under different labels. The specific experimental details are as follows.

\subsection{Dataset}

\begin{figure}[hptb]
	\centering
	\subfigure[]{\includegraphics[width=0.207\textwidth]{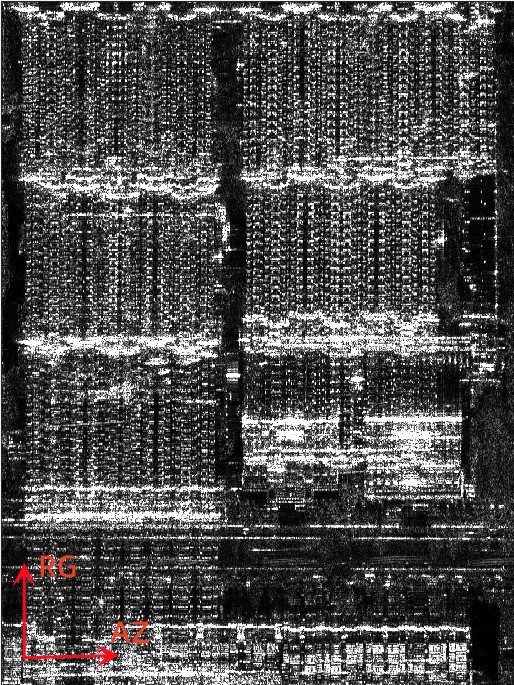}}\vspace*{0em}
	\subfigure[]{\includegraphics[width=0.207\textwidth,height=0.276\textwidth]{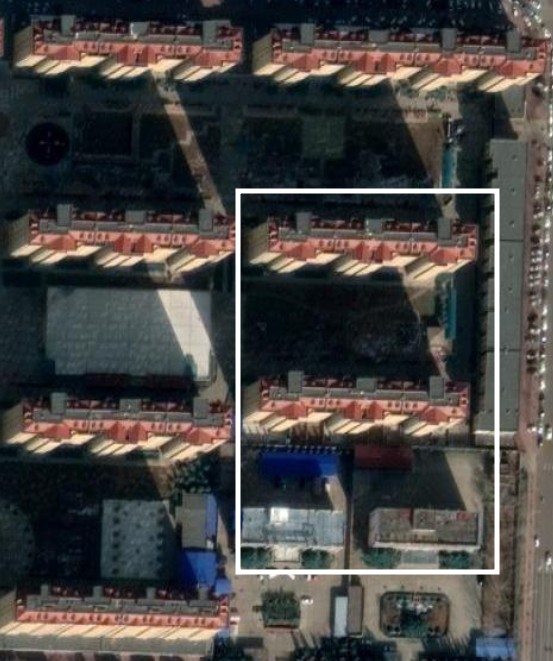}}\vspace*{0em}
	\caption{Real data images of urban area. The area in the white box is the target area of this experiment. (a) SAR image (b) Optical image }
	\label{realdata}
\end{figure}

In this section, we adopt real SAR data from the SARMV3D1.0 dataset~\cite{dataset}. SARMV3D1.0 is an airborne array interferometric SAR system and the data was obtained from an urban community in Wanrong County, Yuncheng City, Shanxi Province, China, by the Aerospace Information Research Institute of the Chinese Academy of Sciences (AIRCAS). The array InSAR system has 8 channels. Table. \ref{parameter} describes the scenario parameters. The optical and SAR images of the area, in which the imaging targets were residential 14-floor buildings, are shown in the white box of Fig. \ref{realdata}. The height of each floor is around 3.5m, so the height of the building is approximately 50 m. The generation of ATASI-Net model in this experiment was based on a previous 3D building simulation experimental data set, which shared the same scene parameters.

\subsection{Comparison of reconstruction performance}

\begin{figure}[hptb]
	\centering
	\subfigure[]{\includegraphics[width=0.38\textwidth]{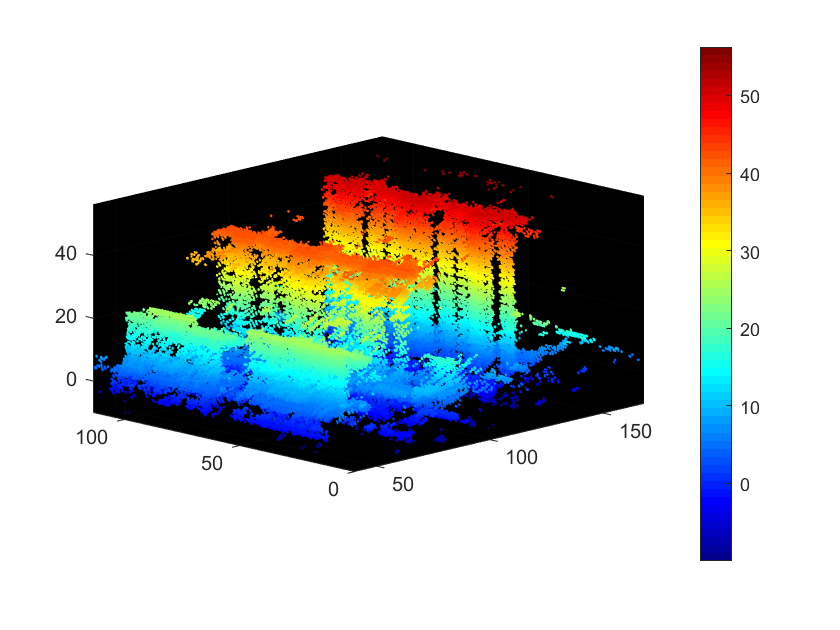}}\vspace*{0em}

	\subfigure[]{\includegraphics[width=0.38\textwidth]{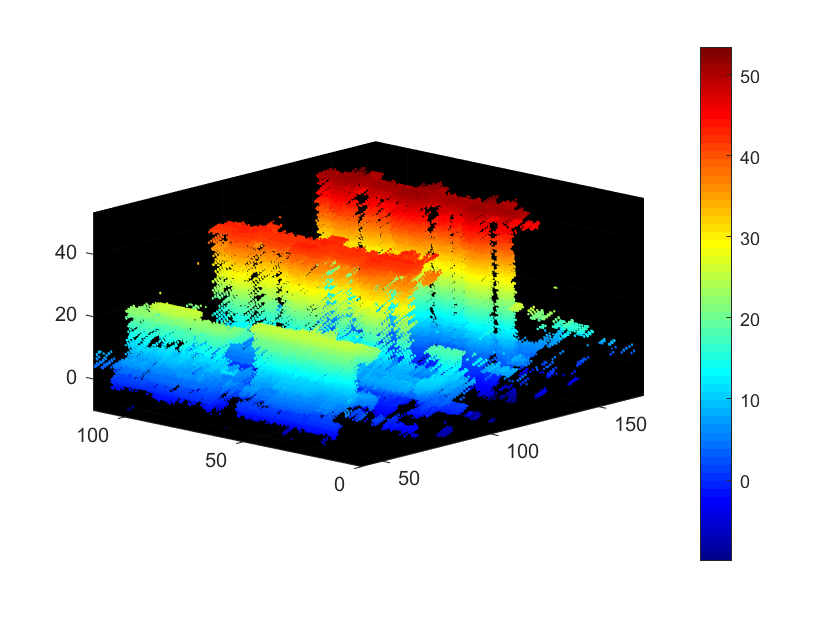}}\vspace*{0em}
	\caption{(a) Point cloud results based on the traditional CS algorithm. (b) Point cloud results based on the proposed ATASI-Net model.}
	\label{ptresult}
\end{figure}

The estimated 3D imaging point clouds leveraging the CS-based method and proposed ATASI-Net are given in Fig. \ref{ptresult}, and Table. \ref{ent1} provides numerical comparisons. Both methods successfully reconstructed the 3D information of the target scene, and the relative heights of the buildings were also accurate. In addition, the reconstruction quality  of the proposed ATASI-Net was substantially greater than that of the traditional CS method, both in terms of the focusing effect of the point clouds and the anti-noise ability. The 3D entropy index in Table. \ref{ent1} can be defined as follows:

\begin{equation}
	\begin{aligned}
	P(i,j) &= \frac{{f(i,j)}}{{({N_{\rm{r}}} \times {N_a} \times {N_z})}}\\
	{\rm{3D Entropy}} &=  - \sum\limits_{i = 0}^{255} {\sum\limits_{j = 0}^{255} {P(i,j) \cdot \ln (P(i,j))} } 
	\end{aligned}
\end{equation}

where $(i,j)$ is the combination of the pixel grey level value $i$ $(0 \le i \le 255)$ and the local mean of neighbor domain $j$ $(0 \le j \le 255)$ to which the pixel belongs. $f(i,j)$ is the statistical quantity of $(i,j)$. And ${N_r},{N_a},{N_z}$ represent the sampling number in the range direction, azimuth direction, and elevation direction, respectively. Entropy is a statistical measure of randomness which can be used to characterize the texture of an image. Here, entropy was adopted to evaluate the focusing quality of the imaging results, with a smaller entropy score indicating higher image quality.

\begin{table}[]
	\caption{Comparisons of the 3D Entropy and Runtime (s) of the Experiments using Real SAR data}
	\label{ent1}
	\centering
	\renewcommand{\arraystretch}{1.7}
	\begin{tabular}{ccc}
		\hline
		\textbf{Method}      & \textbf{3D Entropy} & \textbf{Runtime (s)} \\ \hline
		Traditional CS-based & 0.3780              & 68.964               \\
		Proposed ATASI-Net   & 0.2204              & 0.813                \\ \hline
	\end{tabular}
\end{table}

As shown in Tab. V, ATASI-Net yields lower 3D entropy scores in comparison with traditional CS-based methods, which also indicates that the proposed ATASI-Net is capable of precisely reconstructing images with less clutter and higher focusing quality. In addition, ATASI-Net runs much faster compared with conventional CS-driven algorithms, because the pre-training processing shifts the computational burden to the parameter learning stage, and consequently results in high reconstruction efficiency.

\begin{figure*}[hptb]
	\centering
	\subfigure[]{\includegraphics[width=0.31\textwidth]{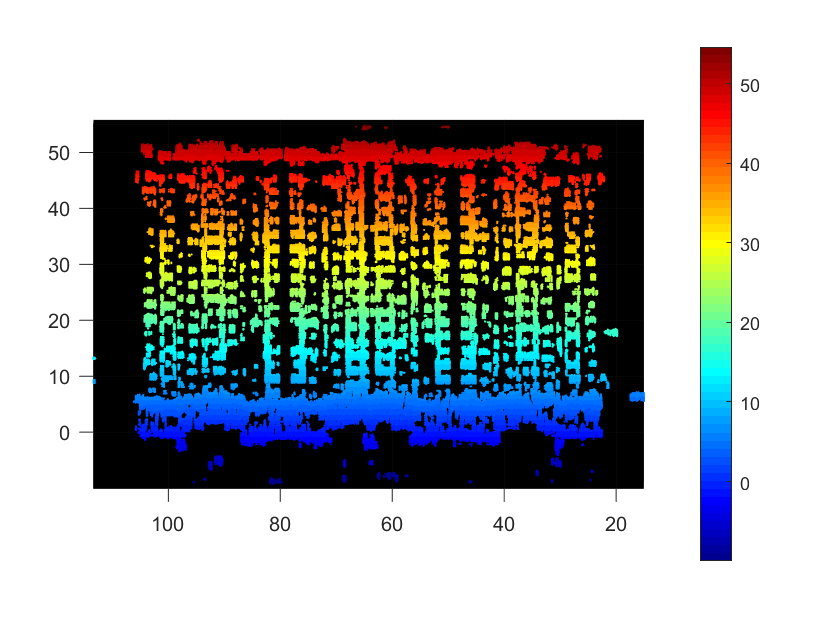}}\vspace*{0em}
	\subfigure[]{\includegraphics[width=0.31\textwidth]{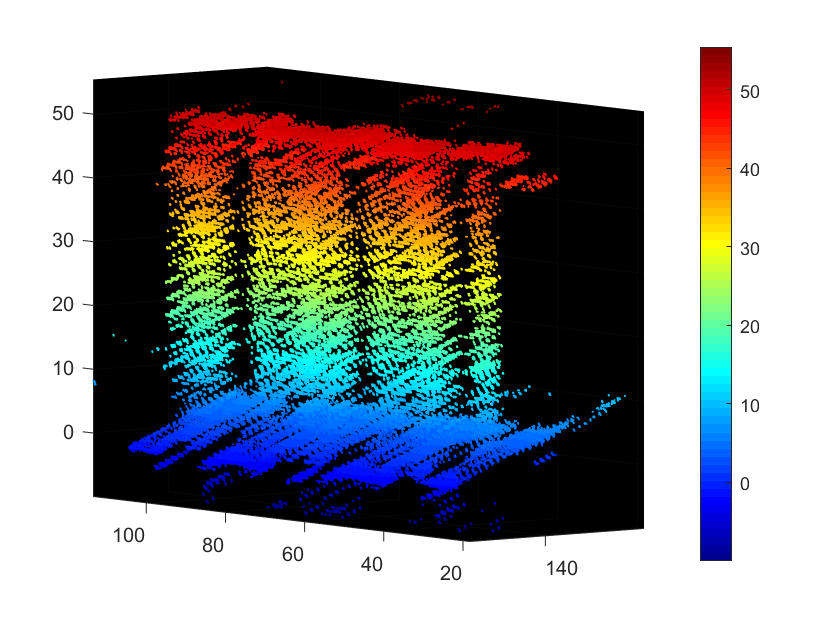}}\vspace*{0em}
	\subfigure[]{\includegraphics[width=0.18\textwidth]{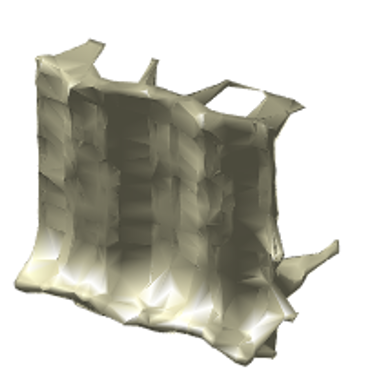}}\vspace*{0em}
	\subfigure[]{\includegraphics[width=0.31\textwidth]{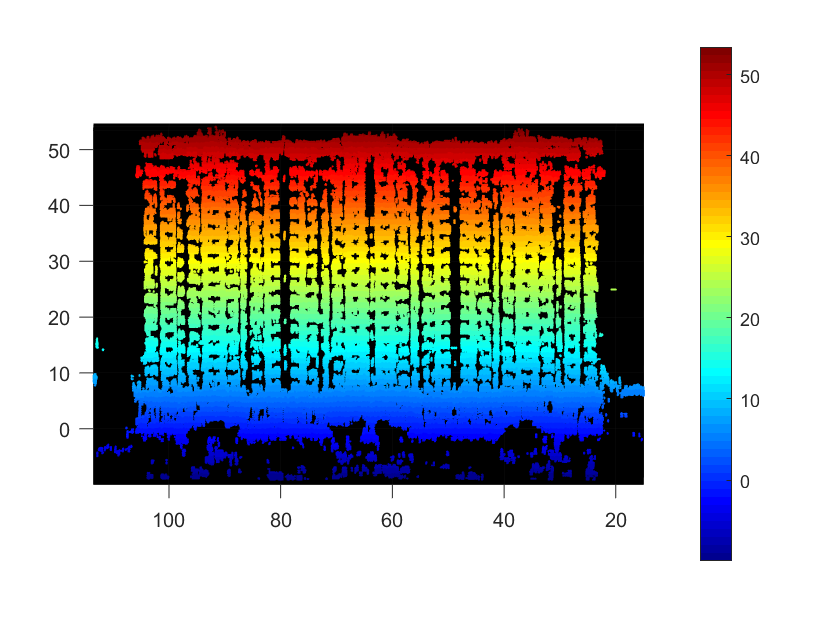}}\vspace*{0em}
	\subfigure[]{\includegraphics[width=0.31\textwidth]{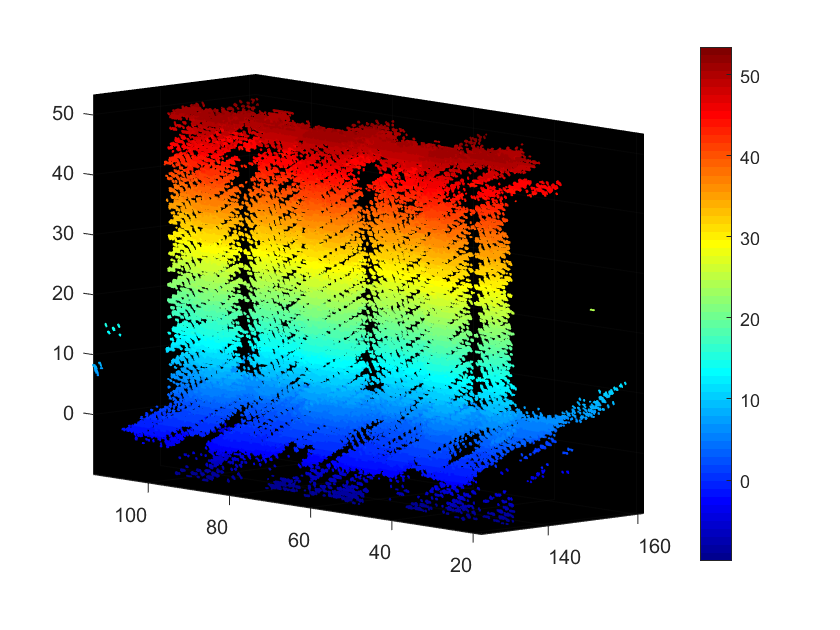}}\vspace*{0em}
	\subfigure[]{\includegraphics[width=0.18\textwidth]{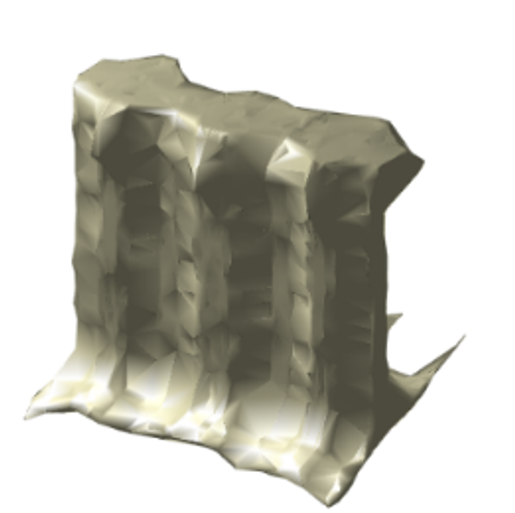}}\vspace*{0em}
	\caption{The point cloud reconstruction results of the last building were selected for structural modeling. (a)(d) The last building point cloud results in front view. (b)(e) The last building point cloud results in side view. (c) Structured modeling of the point cloud results based on the traditional CS algorithm. (f) Structured modeling of the point cloud results based on proposed ATASI-Net. }
	\label{structure-model}
\end{figure*}

The generated point cloud data are often applied to the structural modeling of engineering. To further illustrate the reconstruction effect of the two points clouds, we adopted an alpha-shape-based~\cite{alpha-shape} structural modeling algorithm and selected the last building for structural modeling. The results are displayed in Fig. \ref{structure-model}. The 3D modeling revealed that the point cloud data based on the traditional CS algorithm could not yield a proper model of the building structure, because it has more stray points resulting in poor focus, and the roof and ground cannot be reconstructed well. Whereas, the proposed ATASI-Net successfully attained the 3D modeling of the roof as well as the concave and convex prism of the building.

\subsection{The Ground layer of the traditional CS-based method and the proposed ATASI-Net model}

\begin{figure*}[hptb]
	\centering
	\subfigure[]{\includegraphics[width=0.189\textwidth,height=0.23\textwidth]{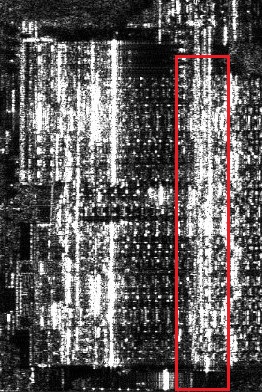}}\vspace*{0em}
	\subfigure[]{\includegraphics[width=0.1665\textwidth,height=0.23\textwidth]{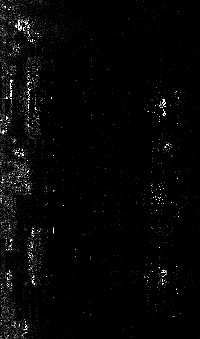}}\vspace*{0em}
	\subfigure[]{\includegraphics[width=0.18\textwidth,height=0.23\textwidth]{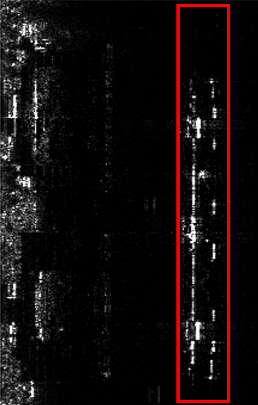}}\vspace*{0em}
	\caption{Slices of ground target reconstruction results. (a) Original SAR image. (b) Traditional CS-based algorithm reconstructs result slice. (c) Proposed method reconstructs result slice. The proposed method has a superior reconstruction capability for weak targets. The ground targets in the red box can be seen in the optical map as fences and cars, for which the proposed method has better reconstruction.}
	\label{qiepian}
\end{figure*}

To further illustrate the performance of the proposed algorithm, we sliced the scattering information of the ground portion of the reconstruction, and the results are shown in Fig. \ref{qiepian}. The area shown in the red box corresponds to the ground information around the building, such as a fence or car, which was not recovered adequately by the traditional CS method but was observed in optical images in Fig. \ref{realdata}. However, the proposed method was capable of retaining the information of some weak objects on the ground because of its element-wise adaptive threshold , resulting in enhanced detail recovery ability. In addition, the focusing effect and reconstruction ability of ATASI-Net were also greater  than those of the traditional CS algorithm.

\subsection{Training labels}

Considering the situation that the imaging parameters of the system cannot be obtained or lost, we hope to realize that it is not necessary to know the parameters of the real imaging scene. Moreover, it is not necessary to generate the training set through simulation first and then use this training set for network training. Therefore, based on the results of the traditional CS method, we used some of the target points of the CS reconstruction as the training labels, without first establishing the simulation scene. The selected area exhibited a more accurate reconstruction effect and fewer stray points, and the chosen training set was the last building shown in Fig. 16 (b). As expected, using the reconstruction obtained with the traditional CS algorithm as the training label improved the result obtained, resulting in fewer stray points and better focus. Thus, by using the result obtained with the traditional method as the training label, the step of reconstructing the simulation scene can be omitted, and the 3D reconstruction process is simplified to a certain extent.

\begin{figure}[hptb]
	\centering
	\subfigure[]{\includegraphics[width=0.38\textwidth]{omp2.png}}\vspace*{0em}
	\subfigure[]{\includegraphics[width=0.38\textwidth]{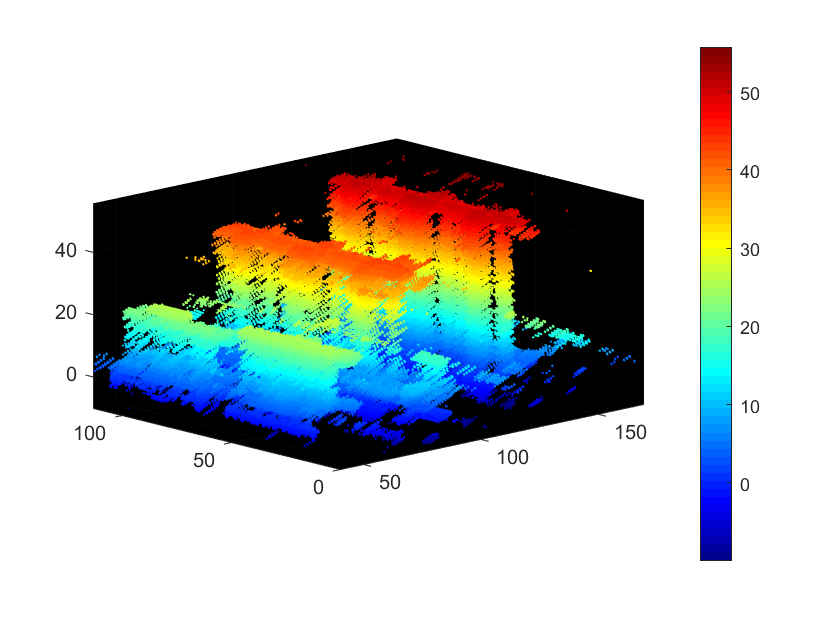}}\vspace*{0em}
	\caption{Comparison of reconstruction results of different training tags. (a) The result of traditional CS algorithm reconstruction, the chosen training set was the area in the last building. (b) Reconstruction result with the selected area based on (a) used as training label. Using the reconstruction obtained with the traditional CS algorithm as the training label improved the result obtained.}
	\label{differ-label}
\end{figure}

\subsection{Threshold-semantic feedbacks}

\begin{figure*}[h]
	\centering
	\subfigure[]{\includegraphics[width=0.3\textwidth]{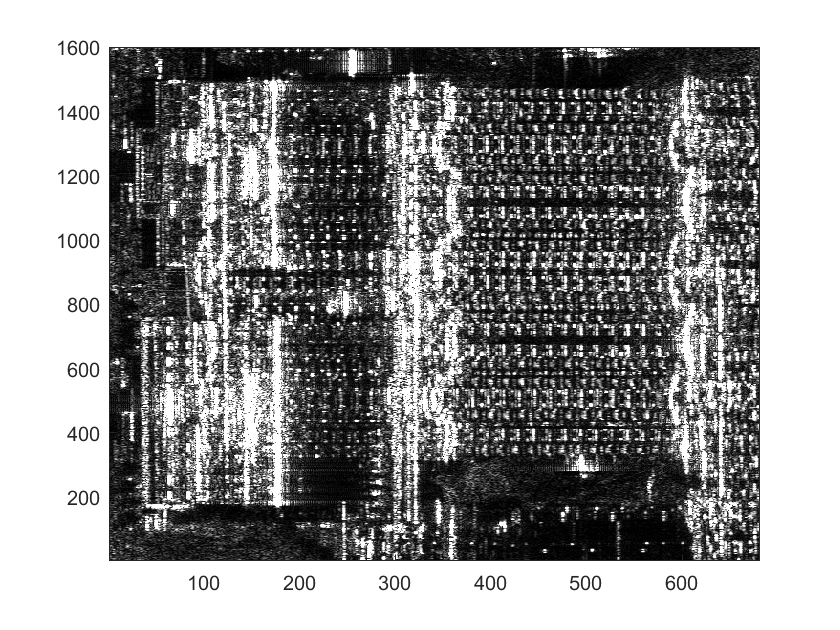}}\vspace*{0em}
	\subfigure[]{\includegraphics[width=0.3\textwidth]{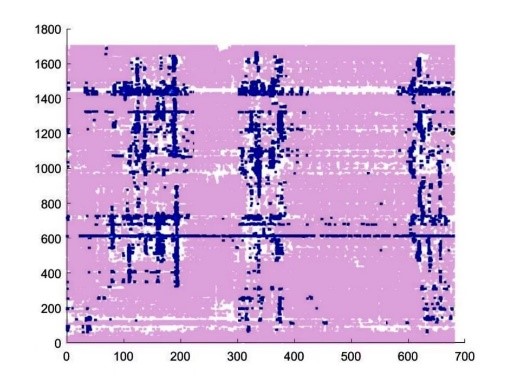}}\vspace*{0em}
	\subfigure[]{\includegraphics[width=0.3\textwidth]{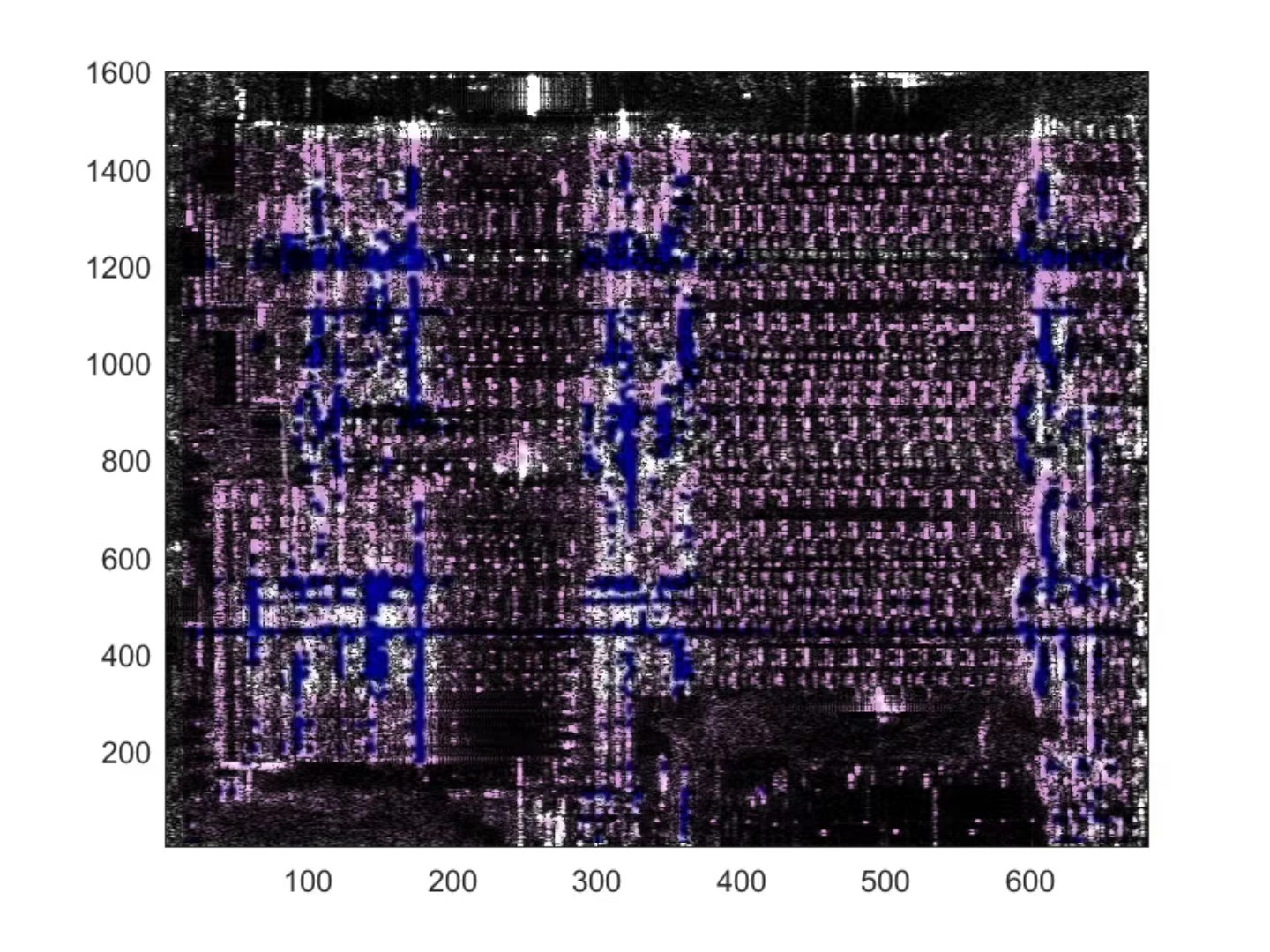}}\vspace*{0em}
	\caption{Segmentation results after threshold visualization. (a) Original SAR image. (b) Threshold segmentation result. (c) The segmentation result fused with the original SAR image. The result of the threshold segmentation is roughly consistent with the actual semantic meaning, which encourages the introduction of semantics as prior constraints. }
	\label{yuyi}
\end{figure*}

Since the proposed method has the characteristic of element-wise adaptive threshold, the threshold learned by the proposed method was visualized, and the threshold was divided into three categories according to the size. This threshold was rendered with different colors, and the results are shown in Fig. \ref{yuyi}. Pink, blue and white represent the background, roof and facade respectively. The classification result of the threshold segmentation is roughly consistent with the actual semantic meaning and exhibits an adequate mutual feed with the semantic meaning. Therefore, in future studies , we will consider introducing semantics into 3D reconstruction as a prior condition to achieve more accurate 3D reconstructions and more effective restoration of information such as trees and other targets that are difficult to distinguish around the building.

\section{Conclusion}

In this article, we propose a novel efficient deep unfolded network, named ATASI-Net, for solving the TomoSAR inversion. The network is designed as a combination of the CS-based iterative algorithm with a data-driven deep learning method, and the parameters are optimized by end-to-end training, thus avoiding complex tuning. The architecture of ATASI-Net is constructed using  the iterative shrinkage-thresholding solver. In addition, the pre-computation module replaces the learning of matrix parameters by solving a convex optimization problem, so that only scalar parameters are used for learning, substantially improving the training efficiency of the network. Moreover, the threshold update unit in ATASI-Net introduces adaptive thresholds, making them not only layer-varied but element-wise. Extensive experiments, including scattering point simulation experiments, 3D building simulation experiments, and a SARMV3D1.0 real dataset experiment, demonstrate the superiority  of the proposed ATASI-Net compared with the conventional CS-based algorithm and the deep-network-based method. Moreover, for real scenes with conventional reconstruction results, we propose using the data showing greater focus as the training set because this enhances the reconstructions in the absence of geometric parameters from real scenes. In addition, we compare the results of the point cloud reconstruction, using structured modeling, which is intuitive and can easily post-process the point clouds. At the same time, we visualize the adaptive thresholds, and the results obtained are able to generate feedback in combination with semantics , which provides practical support for the introduction of semantics as prior constraints to achieve better reconstruction performance in the future.

\end{document}